\documentclass{jfm}
\usepackage{graphicx}
\usepackage{xcolor}
\usepackage{epstopdf, epsfig}
\usepackage{amsmath}
\usepackage{blindtext}
\usepackage{mathtools}
\usepackage[colorinlistoftodos]{todonotes}
\usepackage{placeins}
\usepackage{subcaption}
\usepackage{amssymb}
\usepackage{tikz}
\usepackage{framed}
\usepackage{caption}
\usepackage{enumerate}
\usepackage{siunitx}
\usepackage{url}
\usepackage[colorlinks=true,allcolors=blue]{hyperref}
\usepackage{etoolbox}
\makeatletter
\patchcmd{\@maketitle}{(Received xx; revised xx; accepted xx)\hfill}{}{}%
  {\PackageError{arxiv}{Could not remove the JFM history placeholder}{Check jfm.cls.}}
\makeatother

\shorttitle{Dynamic wetting in the curtain coating}
\shortauthor{Y. Kulkarni, T. Fullana, S. Popinet and S. Zaleski}

\title{
The rapidly advancing contact line Part-1:\\ Navier slip and microscale inertial effects
}

\author{Yash Kulkarni\corresp{\email{\textcolor{blue}{kulkarniyash2398@gmail.com}}}
 \aff{1},  Tomas Fullana\aff{1}, Stéphane Popinet\aff{1} \and Stéphane Zaleski\corresp{\email{\textcolor{blue}{stephane.zaleski@sorbonne-universite.fr}}}\aff{1,2}
}

\affiliation{
\aff{1}Sorbonne Université and CNRS, UMR 7190, Institut Jean le Rond $\partial$’Alembert, F75005 Paris, France

\aff{2}Institut Universitaire de France, UMR 7190, Institut Jean le Rond $\partial$’Alembert, F75005 Paris, France
}

\newcommand{\revised}[1]{{#1}}
\newcommand{\Ca}{\operatorname{Ca}}
\newcommand{\Reyn}{\operatorname{Re}}

\begin{document}
\raggedbottom

\maketitle

\begin{abstract}
Curtain coating, in which a moving plate is coated by a falling liquid sheet, sustains rapidly advancing contact lines at large capillary numbers $\Ca \sim \mathcal{O}(1)$. 
Steady-state solutions exist up to a critical capillary number, beyond which wetting failure occurs through the entrainment of air bubbles. In the steady regime, experiments report an acceleration of the velocity along the fluid-fluid interface as the contact line is approached, down to a few tens of micrometres, and this observation has been interpreted as evidence against the Navier slip model. Here we show that the observed acceleration is compatible with slip models. We perform two-phase Navier--Stokes Volume-of-Fluid simulations with quadtree adaptive mesh refinement using Navier slip and a fixed contact angle. 
Although the velocity vanishes at the contact line in the contact-line frame, the experimentally accessible microscale region lies outside the slip region. The local Reynolds number based on the distance from the contact line is order unity in this region, and therefore local inertial effects cannot be neglected. The scaled interfacial velocity is well described by an inertially corrected wedge flow solution whose wedge angle can be chosen as the interfacial angle at the inflection point, with agreement improving as the slip length is reduced. 
Reducing the slip length also recovers a non-monotonic dependence of the critical capillary number on the feed-flow Reynolds number, beyond which wetting failure occurs. At larger scales, a local Young--Laplace arc describes a finite sessile-drop-like region of the outer heel; this region contracts as the heel is suppressed. 
% The observed interfacial acceleration therefore does not by itself invalidate slip models for rapidly advancing contact lines.
\end{abstract}

\begin{keywords}
contact lines, Navier slip, inertial effects
\end{keywords}

\section{Introduction}\label{sec:Introduction}

A three-phase contact line is the line where a fluid--fluid interface meets a solid surface. The static equilibrium contact line is understood thermodynamically through the Young relation, which has been known for more than two centuries \citep{Young_original}. The dynamics of a moving contact line, however, have remained controversial for decades \citep{Hocking_drop_spreading,Blake_Yulli_IFT,Eggers_comment,Yulli_review}. In this paper, we examine the widely used Navier-slip boundary condition in the curtain-coating configuration, restricting attention to steady two-phase motion in the continuum sharp-interface limit.

Theories of the contact line can be distinguished as to whether they predict the contact angle as a boundary condition
for a continuously differentiable interface, or whether the contact angle is defined at some defined scale, as an angle at some distance $l_M$. In the second case to make the contact angle observable by optical methods
one must specify $l_M \simeq 1 \mu$m or larger. It is clear that the first option leads to a description of the problem
{\em ab initio} while the second description must be inferred from a fundamental theory. We typically call the first angle the nanoscopic contact angle $\theta_n$ while the angle at the scale $l_M$ can be called the apparent contact angle
$\theta_M$. While the apparent angle may be a function of non-local properties of the flow, such as entry flow conditions far from the contact line (see \cite{Blake_1999} for an example), the nanoscopic angle is most often assumed to be a ``local" function of the flow properties. What local means here is sometimes difficult to express precisely, so giving some specific formulations is preferable.
The simplest assumption is a fixed nanoscopic dynamic angle equal to the equilibrium angle
\begin{equation}
    \theta_n = \theta_e .\label{angle}
\end{equation}
Based on kinetic theory considerations and molecular dynamics a more general formulation is obtained \citep{Blake_Haynes, TZ_Qian_GNBC} as

\begin{equation}
    U_{CL} = f(\theta_n) \label{mobrel},
\end{equation}
where the function $f$ depends on the three materials being used. %
In addition for given materials a dependency on local temperature, pressure and electromagnetic field could appear.
The mobility relation (\ref{mobrel}) needs to be associated with a boundary condition for the fluid momentum equations.
For a moving fluid interface, the traditional no-slip boundary condition was shown to have a force singularity at the contact line by \citet{Huh_Scriven} suggesting that the contact line motion is impossible. This was analysed from a Lagrangian point of view by \citet{Dussan_Davis} who showed that under the assumption of the experimentally observed rolling motion for the fluid-fluid interface along with the no-slip boundary condition, the velocity of the contact line is multi-valued. A suggested remedy was to allow a small slip velocity at the fluid-solid interface \citep{Dussan_slip_model}, quantified by the slip length. A simple example is
\begin{equation}
u - \lambda \frac{\partial u}{\partial y} = U_{S}.
\label{eq:NBC_theory}
\end{equation}
Here $u$ represents the tangential component (along the solid substrate) of the fluid velocity at the fluid-solid interface (no-penetration condition sets the normal component $v = 0$ at the solid) and $U_{S}$ represents the velocity of the solid substrate, both are in the laboratory frame of reference. $\lambda$ is the slip length, typically of the order of nanometers. This is known as the \citet{Navier_slip_OLD} slip boundary condition (NBC).

For steady-state contact line motion in the Stokes limit, it was shown by \citet{Hocking_2} that the NBC (constant slip length) removes the force singularity and a finite expression for the wall shear stress, assuming a flat interface was also obtained. The velocity $u$ at the contact line is also related to the contact line velocity by
\begin{equation}
U_{CL} = u(x_{CL}) - U_{S} ,
\end{equation}
where $x_{CL}$ is the contact line position.
A more general class of mobility relation can occur if the local properties of the flow field are taken into account.
For example, the Generalized Navier Boundary Condition for a sharp interface method \citep{TZ_Qian_GNBC} reads
\begin{equation}
U_{CL} - \lambda \partial_y u  = \frac{\sigma}{{\mu_l}}  [ \cos(\theta) - \cos(\theta_s) ] g ( \frac x \epsilon) ,
\end{equation}
where $\theta_s$ is the static contact angle, ${\mu_l}$ the liquid viscosity, $g$ a smoothed $\delta$ function and $\epsilon$ a microscopic length.
Assuming a $C^1$ velocity field up to the contact line, the normal vector at the interface (and hence the contact angle) rotates as the contact line moves \citep{Fricke_paradox}, resulting in $\partial_y u = - \dot \theta$ and yielding the
generalized mobility law
\begin{equation}
U_{CL} = f(\theta, \dot \theta ) = f(\theta, - \partial_y u) .
\end{equation}
Such a law is still local, albeit more general than the elementary mobility law (\ref{mobrel}).
Even more general laws may be found in the form
\begin{equation}
U_{CL} = f( \theta, \dot \theta , \chi_1, \chi_2),
\end{equation}
where $\chi_1$, $\chi_2$ are some local properties of the flow, such as derivatives of the velocity field or of the interface height. For example, the super slip condition involves $\partial^2_{yy} u$ \citep{Hocking_YK}.
Many other boundary conditions have been suggested \citep{Bonn_Eggers_Rev} sometimes involving additional physical effects (intermolecular forces, tensio-active materials, interface roughness, non-constant surface tension \cite{IFM_Yulli}). Almost all of these theories involve very small length scales and are thus extremely difficult to verify experimentally. Moreover it is well known
\citep{Cox_log_law,Voinov} that large curvatures $1/R$ appear near the contact line, so that the angle predicted at the nanometer scale may be very different from the angle observable by optical methods at the micrometer scale. As a result theories of the contact line boundary condition such as the NBC may only be tested indirectly, solving the Navier-Stokes equations over the whole range of scales from below the nanoscopic scales $\lambda$ or $R$ to the outer flow scales.
The problem becomes especially challenging when the contact line advances rapidly. In curtain coating, steady solutions with $\Ca=O(1)$ are possible, so the contact-line speed can be large enough for the local Reynolds number based on the distance from the contact line to become order unity already at experimentally accessible micrometric scales.
In that regime, a purely Stokes-flow description is no longer sufficient in the observable region, so asymptotic theories based on creeping-flow assumptions, such as those of \citet{Cox_log_law,Voinov}, do not directly apply there. Since the interface can bend substantially between the experimentally resolved micrometric scale and the nanometric scale at which a microscopic contact-line model is postulated, resolving the full multiscale structure becomes a formidable challenge.
As far as the authors know the only works that cover the full range of scales from the nanoscopic to the centimetre scale are those of \citet{Wilson_Yulli} in the curtain-coating setup and \cite{Kamal_Eggers_cusp} in the cusp setup. Both of these works assumed a free surface. While \citet{Wilson_Yulli} solve the full Navier-Stokes, \citet{Kamal_Eggers_Cox_Voinov} solve the Stokes flow only. This paper aims to do a full-range computation with the two-phase Navier-Stokes equations. However, it turns out that a simulation that starts with a grid size that accurately resolves the nanometer scale, and hence is markedly smaller than the nanometer, and reaches the scale of several centimeters is out of reach with current Volume-Of-Fluid methods or any other two-phase direct numerical simulation methods. This paper thus explores a very wide, but not the full range of scales of a typical experiment, going down only to slip lengths of a hundred nanometers.

\subsection{The curtain coating setup}
\begin{figure}
  \centerline{\includegraphics[width=\textwidth]{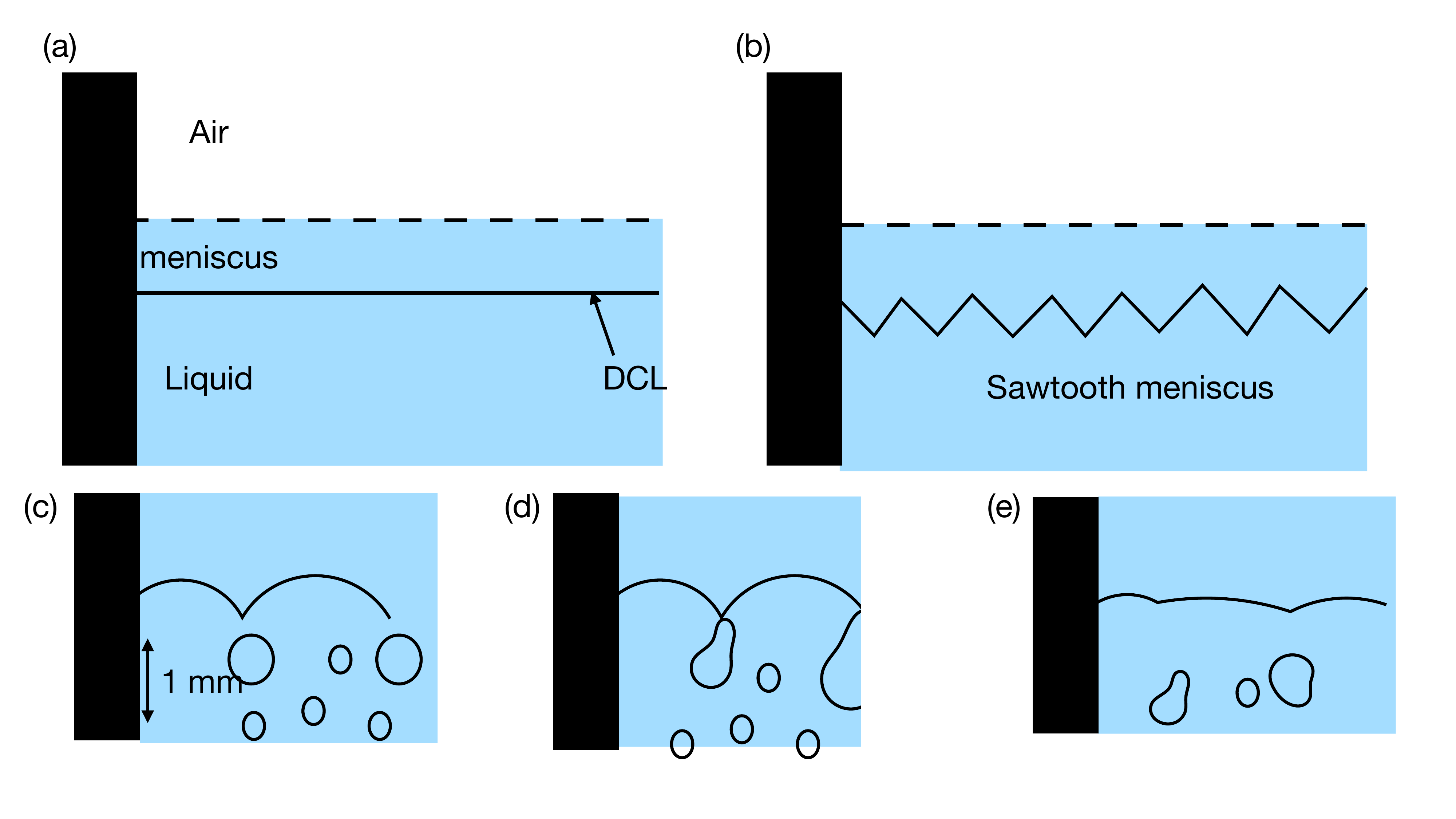}}
   \caption{Schematic representation of air entrainment and wetting failure, based on the experimental sequence shown in Fig.~4 of \citet{Vandre_exp}.
(a) Stable steady-state condition.
(b) Onset of wetting failure, marked by the appearance of a sawtooth meniscus.
(c) Entrainment of an air film.
(d) Rupture of the air film and bubble formation.
(e) Bubbles and residual air films left on the coated substrate.}
\label{wetting_failure_exp}
\end{figure}

The problem we investigate numerically in this paper is curtain coating, which is a dynamic wetting setup. In such setups, a liquid
displaces a gas (usually ambient air) on a flat solid substrate or in other words, coats the solid substrate. The dynamic wetting phenomenon has wide industrial applications as it forms the basis of many coating processes. It is known that beyond a critical substrate speed, wetting failure occurs, which is marked by air entrainment in the form of thin films and bubbles. Figure \ref{wetting_failure_exp}, taken from the experiments of \citet{Vandre_exp}, shows the onset of wetting failure for the plunging plate setup. Such entrained bubbles and films degrade the quality of the coated product and are often undesirable \citep{coating_applications}. Thus, one would like to delay wetting failure as much as possible in industrial applications. The {\em hydrodynamic assist} effect below indeed speeds up to the order of meters per second \citep{Blake_1999}.

\begin{figure}
  \centerline{\includegraphics[width=\textwidth]{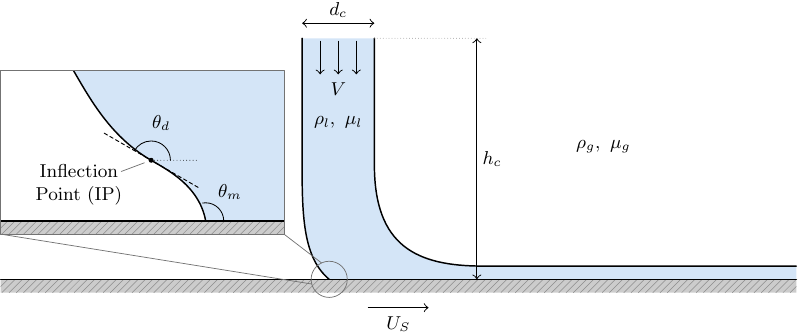}}

  \caption{Schematic of the curtain coating configuration. The system parameters are $h_c$ the curtain height, $d_c$ the curtain width, $\rho_l$, $\rho_g$ and $\mu_l$, $\mu_g$ the densities and viscosities of the liquid phase and the gas phase respectively, $U_{S}$ the substrate velocity, $V$ the feed flow velocity and $\theta_m$ the imposed contact angle. The inflection point IP corresponds to the point at which the curvature of the interface is zero and the angle measured here is denoted by $\theta_d$.}
\label{curtain_Tomas}
\end{figure}

The schematic of the curtain coating setup is shown in Figure \ref{curtain_Tomas}. The liquid is falling from the top with a feed flow velocity $V$, the substrate is being pulled with a velocity $U_{S}$, and a liquid film with constant thickness coats the substrate. An interesting property of this setup is that by altering the feed flow velocity $V$ one can increase the critical value of $U_{S}$ (beyond which wetting failure occurs) \citep{Blake_1999}. This is the `hydrodynamic assist' effect. The increase in the inertia of the impinging curtain causes an increase in the pressure in the liquid in the vicinity of the contact line which `assists' in pushing the air away from the contact line  \citep{Liu16}. %

\begin{figure}
\centering
\includegraphics[width=\textwidth]{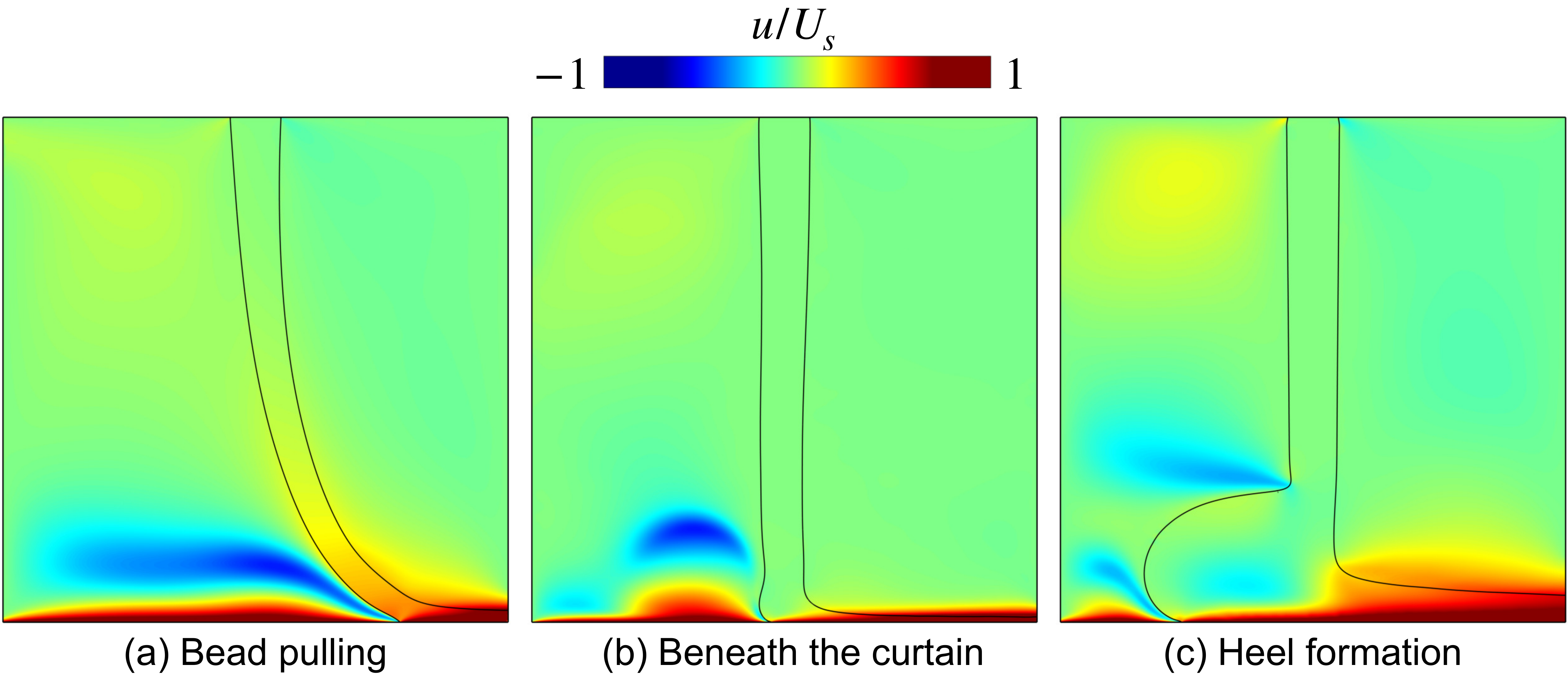}
\caption{Representative steady curtain-coating configurations obtained in the reduced model: (a) bead pulling, $\Reyn=11$ and $\Ca=0.5$; (b) beneath-the-curtain configuration, $\Reyn=20$ and $\Ca=1.8$; and (c) heel formation, $\Reyn=30$ and $\Ca=0.7$. The notation and parameter choices follow Figure~\ref{curtain_Tomas}: the curtain height is $h_c=1\,\mathrm{cm}$, which is also the domain height, and the curtain width is $d_c=1\,\mathrm{mm}$. The full computational domain is shown in each panel, with slip length $\lambda=10\,\mu\mathrm{m}$. The background is colored by $u/U_{S}$, where $u$ is the horizontal component of the velocity and $U_{S}$ is the solid-plate velocity, directed from left to right in the image. The Reynolds number $\Reyn$ is based on the vertical feed-flow velocity $V$, and the capillary number $\Ca$ is based on the horizontal plate velocity $U_{S}$.}
\label{fig:curtain_config_bview}
\end{figure}

In the steady regime, the large-scale shape of the liquid curtain near the contact line can take three representative forms \citep{Karim_curtain_non_newtonian_1}: bead pulling, in which the contact line lies downstream of the falling curtain; a beneath-the-curtain configuration, in which the contact line remains below the impinging liquid sheet; and heel formation, in which the contact line lies upstream of the curtain.
Figure~\ref{fig:curtain_config_bview} shows these three representative configurations observed in the simulations. These configurations provide a useful macroscopic description of the steady solutions before the onset of wetting failure. In general, larger feed-flow velocity favours heel formation, whereas larger plate speed favours bead pulling. The beneath-the-curtain configuration occurs for intermediate combinations of feed-flow and plate speed.

Wetting failure originates from the instability of the contact line and hence, the contact line dynamics are crucial to predicting the correct critical speed at the onset of wetting failure. As any fully resolved no-slip model can never predict any stable solutions of the dynamic contact line \footnote{In VoF formulation, there is an implicit numerical slip, scaling with the grid size that removes the no-slip paradox and steady solutions are possible \citep{Afkhami_SZ}} (due to the discontinuity of the contact line velocity), the stable solutions are often studied numerically by the slip models along with an imposed constant microscopic contact angle (\citet{Kamal_Eggers_Cox_Voinov}, \citet{Kamal_Eggers_cusp}, \citet{Afkhami_SZ}, \citet{Liu16}). This procedure (use of slip model) was questioned by \citet{Wilson_Yulli}, \citet{Yulli_review} who performed simulations of the curtain coating setup using the slip boundary condition and neglecting the air phase. They tried to compare their numerical results against the experiments of \citet{Blake_1999} and not only did the simulations fail to predict any critical wetting speed, but the variation in the dynamic or apparent contact angle was significantly narrower than observed in the experiments. Because of this, \citet{Wilson_Yulli} concluded that the microscopic angle is also a function of macroscopic flow parameters like the Reynolds number based on the feed velocity. Later, \citet{Liu16} demonstrated that assuming a constant contact angle at the grid scale and a slip boundary condition, while considering air stresses, one can predict some critical capillary number beyond which wetting failure happens.
This would not have been possible with a free surface, where all capillary numbers result in a stable steady-state solution. \citet{Liu16} tried this for a reduced setup with a ten-micron slip length and not for the parameters corresponding to the Blake experiments, which require nanometric slip lengths and hence huge computational costs.

However, the flow structure in the immediate vicinity of the contact line remains unclear. Experiments \citep{Clarke_black_white} show that the velocity along the interface increases as the contact line is approached down to the smallest observable scales (of order $20\,\mu$m). In contrast, viscous wedge solutions with slip predict a decrease of velocity towards the contact line \citep{Hocking_2}. This discrepancy has been interpreted as evidence against slip-type models \citep{Yulli_review}.
At the same time, the inertially corrected wedge flow analysis of \citet{Varma} predicts an increase of interfacial velocity when inertial effects are significant. In curtain coating, the local Reynolds number based on the distance from the contact line becomes order unity at micrometric scales, suggesting that inertia may control the observable flow even though the microscopic regularisation occurs at nanometric scales. Furthermore, different microscopic regularisations are known to produce nearly identical meniscus shapes at macroscopic distances \citep{Dussan_slip_model}.
The present work investigates this possibility using direct numerical simulations of the two-phase Navier–Stokes equations with a Navier-slip boundary condition. Our objective is to determine which flow region selects the observable interface shape and velocity field, and how this connects microscopic regularisation to macroscopic behaviour. Our results show that qualitative macroscopic observations in rapidly advancing contact lines do not provide a decisive falsification of slip-type boundary conditions because the observable flow is controlled by hydrodynamic structure occurring outside the microscopic regularisation region.
In other words, the supposed falsification is not decisive in this setup.

\subsection{Overview of this paper}
We first present the governing equations and numerical method used to simulate the two-phase curtain-coating configuration with a Navier-slip boundary condition.
We then verify the numerical model by reproducing the reduced curtain-coating results of \citet{Liu16} and by demonstrating convergence of the solution down to the slip-length scale, including numerical evidence of the logarithmic curvature singularity at the contact line \citep{Hocking_YK,Olivier_SZ}.
Next, we characterise the apparent contact angle measured at the inflection point located at micrometric distances from the contact line. We then analyse the velocity field along the interface and show that the experimentally observed acceleration near the contact line arises from an inertial region where the local Reynolds number is of order unity. The flow in this region is compared with the inertially corrected wedge solution of \citet{Varma}, where the wedge angle is taken as the angle at the inflection point, and agreement improves as the slip length is reduced towards sub-micrometric values.
\revised{Finally, we test a sessile-drop-like outer shape, motivated by the reduced speed within the heel. Using measured turning-point data, we construct a local Young--Laplace arc and assess it through both interface position and a tangent-based diagnostic. We propose this as a candidate outer shape to which the IC-SFW region may connect.} Concluding remarks and perspectives are presented in \S\ref{sec:conclusion}.
\section{Methodology} \label{Methodology}

\subsection{Mathematical model}

The mathematical model is based on the mass and momentum conservation equations for incompressible and isothermal flow with variable density and surface tension force resulting in

\begin{equation}\label{eq:mass}
\nabla\cdot\vec{u} = 0 ,
\end{equation}
\begin{equation}\label{eq:NS-continuum}
\frac{\partial \rho \vec{u}}{\partial t} + \nabla\cdot(\rho\vec{u}\vec{u})
=
- \nabla p
+ \nabla\cdot \left(2\mu\mathbf{D}\right)
+ {\rho}\vec{g}
+ f_\sigma,
\end{equation}

\begin{equation}\label{eq:density_var}
    \frac{\partial \rho}{\partial t} + \nabla\cdot (\rho \vec{u}) = 0,
\end{equation}

where $\vec{u}(\vec{x},t)$ is the velocity field and $p(\vec{x},t)$ is the pressure field. The tensor $\mathbf{D}$ is defined as $\frac{1}{2}\left[\nabla \vec{u} + \left(\nabla \vec{u}\right)^T\right]$. $\rho$ and $\mu$ are the density and viscosity respectively and $\vec{g}$ is the acceleration due to gravity. $f_{\sigma}$ is the surface tension force defined as

\begin{equation}\label{eq:STF}
f_\sigma= \sigma \kappa \mathbf{\hat{n}} \delta_s (\vec{r} - \vec{r}_f),
\end{equation}

which depends on the surface tension coefficient $\sigma$ and the interface shape $\vec{r_f}$, particularly on its curvature $\kappa$ and normal $\vec{n}_s$. The Dirac function $\delta_s (\vec{r} - \vec{r}_f)$ indicates that the force only acts at the interface and is zero everywhere else.

At the contact line, we assume a constant angle as in equation \ref{angle} and a Navier slip boundary condition for the fluid-solid interface. As mentioned in section \ref{sec:Introduction}, we impose the Navier slip boundary condition (NBC) (\ref{eq:NBC_theory}).

\subsection{Numerical method}

We use the free solver \textit{Basilisk} (\url{https://basilisk.fr/}) developed at our institute by Stephane Popinet and various collaborators (\citet{Basilisk_Popinet_1}, \citet{Basilisk_Popinet_2}, \citet{Basilisk_Popinet_3}, \citet{Afkhami_Bussmann_2D}, \citet{Afkhami_Bussmann_3D}).
The two-phase interfacial flow capabilities of \textit{Basilisk} focusing on the contact lines are well-tested \citep{Basilisk_test_3,Tomas_EPJST_1,Tomas_EPJST_2}.
A detailed description of the adaptive two-phase solver and the height-function surface-tension treatment can be found in \citet{Basilisk_Popinet_1}. Here, we give only a brief description of the ingredients relevant to the present contact-line simulations.
An overview of the temporal and spatial discretisation used by the solver is provided in Appendix~\ref{appA}.

A volume fraction field $c$ is introduced, defined as 1 in fluid `1' (liquid in this case) and 0 in fluid `2' (gas). The interface is tracked by solving the advection equation for the volume fraction

\begin{equation}\label{eq:c-advect}
\frac{\partial c}{\partial t} + \nabla\cdot(c\vec{u})=0.
\end{equation}

The density and viscosity are then defined as
\begin{equation}\label{eq:properties}
\rho  = c \rho_1 + (1 - c)\rho_2.
\qquad
\mu  = c{\mu_l} + (1 - c){\mu_g}.
\end{equation}

The contact angle boundary condition defines the normal vector at the contact line and hence affects the curvature calculation at the boundary and hence the calculation of the surface tension force. The contact angle is imposed at the boundary of the interface using a VoF tracer and the height function field, according to \citet{Afkhami_Bussmann_2D} approach. The reader is directed to \citet{Afkhami_Bussmann_2D} and \citet{Afkhami_Bussmann_3D} for full details. Briefly said, we define a ghost fluid layer below the domain and then set the tangential component of the height function field to give the desired normal vector at the contact line and hence the desired contact angle. The corresponding Basilisk implementation is available in the contact-angle header \url{https://basilisk.fr/src/contact.h}.

\begin{figure}
    \centering
    \includegraphics[width=1\linewidth]{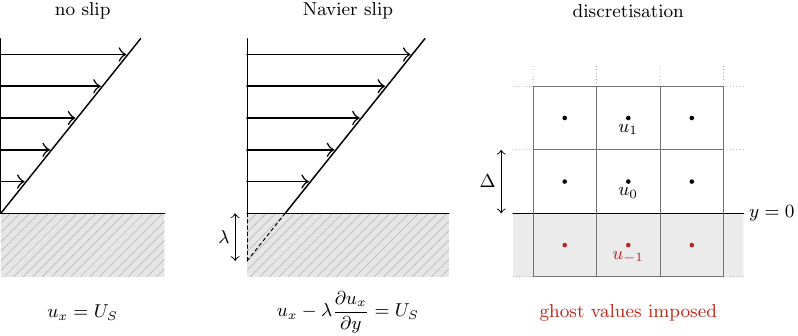}
    \caption{Visualisation of the slip length and numerical discretisation of the Navier slip boundary condition. The grey boxes below the x-axis are the ghost cells where the value is imposed according to equation \eqref{eq:NBC_discrete}. }
\label{fig:NBC_discrete_Tomas}
\end{figure}

The Navier slip condition \eqref{eq:NBC_theory} is imposed through the same layer of ghost cells. Geometrically, $\lambda$ is the distance below the substrate at which the linear extrapolation of the velocity profile reaches the substrate velocity $U_{S}$ (Figure~\ref{fig:NBC_discrete_Tomas}). All variables are collocated at cell centres (Appendix~\ref{appA}). Let $u_0$ denote the tangential velocity in the first fluid cell, centred at $y = \Delta/2$, and $u_{-1}$ its value in the ghost cell beneath it, centred at $y = -\Delta/2$, where $\Delta$ is the local grid size. Linear interpolation between these two values gives the wall velocity and the wall shear rate, both to second order in $\Delta$:
\begin{equation}
u_x\big|_{y=0} \simeq \frac{u_{-1} + u_0}{2} ,
\qquad
\frac{\partial u_x}{\partial y}\bigg|_{y=0} \simeq \frac{u_0 - u_{-1}}{\Delta} .
\label{eq:NBC_interp}
\end{equation}
Substituting \eqref{eq:NBC_interp} into \eqref{eq:NBC_theory} and solving for the ghost-cell value gives
\begin{equation}
\frac{u_{-1} + u_0}{2} - \lambda \, \frac{u_0 - u_{-1}}{\Delta} = U_{S}
\quad \Longleftrightarrow \quad
u_{-1} = \frac{2\Delta}{2\lambda + \Delta} \, U_{S} + \frac{2\lambda - \Delta}{2\lambda + \Delta} \, u_0 .
\label{eq:NBC_discrete}
\end{equation}
For further details on the Navier slip condition implementation and validation within the same framework we refer the reader to \citet{Tomas_EPJST_1, Tomas_EPJST_2}.

\subsection{Simulations setup}\label{subsec:simulation_setup}

The simulation setup is shown in Figure \ref{curtain_Tomas}. We define the Reynolds number, capillary number and Bond number as follows:

\begin{equation}
    \Reyn = \frac{\rho_l V d_c}{\mu_l} ,\hspace{1cm} \Ca = \frac{\mu_l U_{S}}{\sigma} ,\hspace{1cm}  Bo = \left( \frac{\rho_l g}{\sigma} \right) \left( \frac{d_c V}{U_{S}} \right)^2.
\end{equation}

$\sigma$ represents the surface tension and all other parameters are shown in Figure \ref{curtain_Tomas}. The gravity $g$ is always assumed to be $9.81 m/s^2$. To determine if we reach a steady state or not, we calculate the difference between the velocity field, the contact line position and the inflection point position at subsequent timesteps. If the difference for each of them is lower than a threshold, we conclude that a steady state is reached. %
For a given $\Reyn$, we vary the $\Ca$ and predict a critical capillary number ($\Ca_{cr}$) beyond which no steady-state solutions are found. The stability window is then deduced from the plot of the smallest $\Ca_{cr}$ for each $\Reyn$.

To study the effect of slip length, we vary the slip length from $10$ microns to a few hundreds of nanometers. The resolution for the resolved slip length studies is reported in terms of grid size per slip length $\left(\frac{\lambda}{\Delta} \right)$. We reached a maximum refinement level of 17 at our computational limit. For a $1\,\mathrm{cm}$ domain, refinement level 17 gives a minimum cell size of $\Delta=1\,\mathrm{cm}/2^{17}\simeq76.3\,\mathrm{nm}$.

\section{Results}\label{results}

The central issue of this paper is whether the micrometric flow features observed in curtain-coating experiments can be used to assess the validity of slip-type contact-line models. In particular, experiments show an apparent acceleration of the interfacial flow as the contact line is approached, whereas a purely viscous slip-regularised inner solution would predict deceleration inside the slip region \citep{Yulli_review}. In the curtain-coating configuration, however, the local Reynolds number at experimentally accessible distances is already of order unity, suggesting that the observed micrometric region may belong to an inertial intermediate scale rather than to the microscopic regularisation region.

We organise the results as follows. We first analyse the full curtain, then compare the interfacial velocity with the inertially corrected wedge solution and discuss the interface bending at larger scales. We then turn to the reduced configuration, where resolved-slip computations allow convergence studies, verification of the logarithmic singularity, and systematic exploration of the effect of slip-length reduction. Unless otherwise stated, all results use the reference parameter set of \citet{Liu16}, introduced later in \S\ref{subsec:compare_exp_Liu}.

\subsection{The liquid curtain and the contact line flow physics}

We begin with the full curtain-coating configuration in order to examine the flow structure near the advancing contact line in the complete two-phase setting. We first document the apparent acceleration in the full curtain simulations, then quantify it via the interfacial-velocity profiles and compare with the inertially corrected Stokes Flow Wedge solution (IC-SFW). This reveals that the apparent accelerating contact-line flow is an outer inertial structure which is truncated only inside the slip region, allowing the slip model and experimental observations to coexist without contradiction.

In the reference frame of the contact line, it is known \citep{Moffatt_1964} that the Stokes flow wedge solution (SFW) for the three-phase contact line with a no-slip boundary condition gives the streamfunction $\Psi \sim rF(\alpha,\theta)$, where $\alpha$ is the wedge angle, $r$ is the radial coordinate and $\theta$ is the angular coordinate assuming the contact line at the origin. Hence the velocity field depends on the angular coordinate and not on the radial coordinate. As a consequence, the SFW flow field is discontinuous at the contact line. From the analysis of \citet{Hocking_2} and \citet{Hocking_YK} in the flat-interface and liquid-vacuum limit, one can conclude that for the Navier-slip boundary condition, the velocity field varies linearly with the radial coordinate or the distance from the contact line, $u \sim r/\lambda$. This means that the velocity of the fluid decreases as one approaches the contact line and becomes exactly zero at the contact line, which makes the solutions presented by \citet{Hocking_2} and \citet{Dussan_slip_model} for slip models continuous at the contact line.

Any Stokes flow subjected to a Navier-slip-like boundary condition will represent a no-slip-like solution in the far field and a slip model in the inner region. This indicates that the velocity along the interface in such a flow can only decrease up to the contact line. However, the experimental measurements for the curtain-coating setup by \citet{Clarke_black_white} and \citet{Kistler_PhD} show that the velocity along the interface increases up to the minimum resolution of the experiments, which was $20\,\mu$m.

The increase in velocity can be attributed to inertia. We show now that, due to the high capillary number of the curtain-coating setup, inertial effects cannot be neglected even at a few tens of microns. \citet{Varma} solved the inertially corrected Stokes flow in a wedge (IC-SFW) subject to the no-slip boundary condition and found that the velocity field accelerates up to the contact line. In particular, \citet{Varma} showed that the velocity along the interface acquires a dependence on $r$. For an advancing contact line, the velocity magnitude along the interface for IC-SFW is smaller than the constant SFW value at a given angle, but approaches the SFW value at the contact line.

The relevant parameter is the local Reynolds number defined as $\Bar{r}= {\rho_l} |U_{S}| r/{\mu_l}$, where $|U_{S}|$ is the contact-line speed in the reference frame of the substrate, $r$ is the distance of a point in the fluid continuum from the contact line, and ${\rho_l}$ and ${\mu_l}$ are liquid density and viscosity respectively. This $\Bar{r}$ is $\mathcal{O}(1)$ at $9\,\mu$m for the current setup assuming $\Ca=1$. For the experiments of \citet{Clarke_black_white} ($U_{S} \sim 1$--$2\,\mathrm{m\,s^{-1}}$, $\mu_l = 68\,\mathrm{mPa\,s}$ and $\sigma = 61\,\mathrm{mN\,m^{-1}}$), the value of $\Bar{r}$ is of order unity at tens of microns. This indicates that inertia cannot be neglected at the experimental resolution of $20\,\mu$m.

\begin{figure}
\centering
\includegraphics[width=\textwidth]{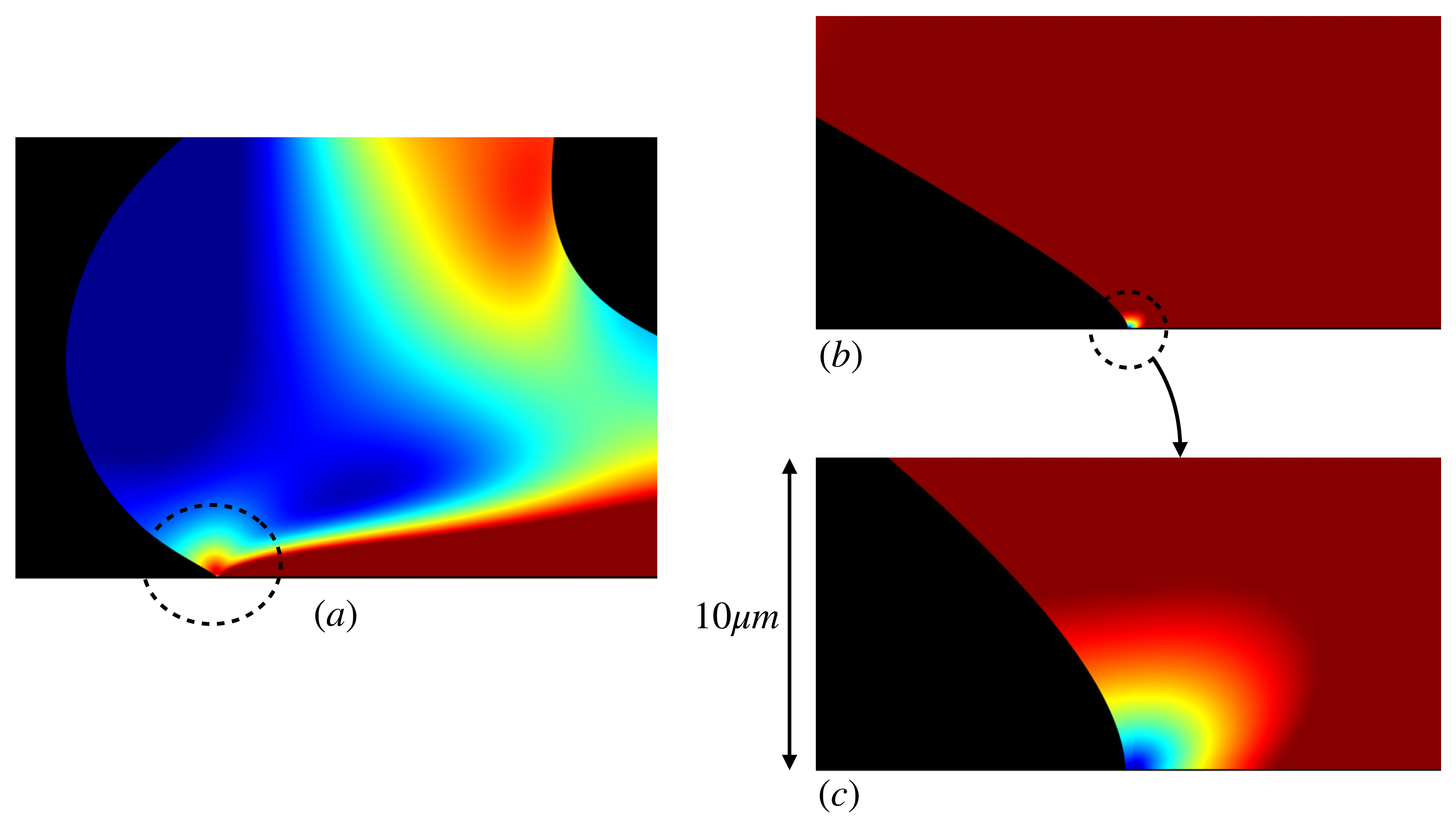}
\caption{Zoom in at the contact line for the steady-state liquid curtain for $\Reyn = 20$ and $\Ca = 0.7$ in the laboratory frame of reference. The solid is moving from left to right and the contact line is at rest. Coloring is done by the magnitude of the velocity such that red is high and blue is low. It can be seen that at apparent scale the flow appears to accelerate as it approaches the contact line (a). Zooming in on the slip region (b,c), we see that the flow decelerates and the velocity approaches zero at the contact line (c). The slip length is $5\,\mu$m.}
\label{fig:speed_Basilisk_contour_iface}
\end{figure}

\begin{figure}
\centering
\includegraphics[width=\textwidth]{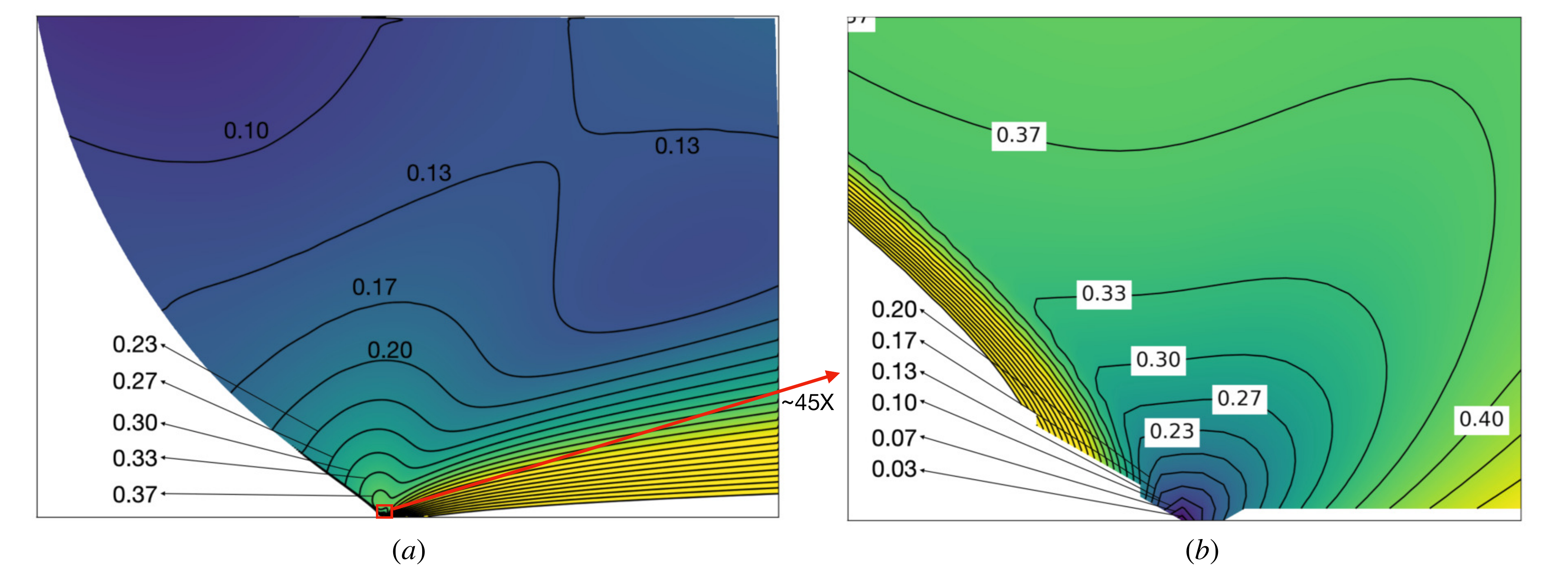}
\caption{Contour plot showing the magnitude of the velocity field for Figure \ref{fig:speed_Basilisk_contour_iface}. The labels indicate the speed magnitude of the contour lines. A fan-like structure is seen in the zoomed-out figure (a), where the magnitude is seen to increase as the contact line is approached. In the zoomed-in image (b), the speed contour values decrease up to the zero velocity at the contact line. The speed labels are scaled by the solid speed, taken to be $1$.}
\label{fig:speed_python_contour_iface}
\end{figure}

Since we solve the two-phase Navier--Stokes equations, such effects are already present in our simulations. Figure \ref{fig:speed_Basilisk_contour_iface} shows the steady-state liquid curtain colored by the velocity magnitude. We see that near the contact line, at apparent scale, the flow accelerates. This is identified by the red fan-like structure around the contact line in Figure \ref{fig:speed_Basilisk_contour_iface}a. When we zoom in at the scale of the slip length, shown in Figure \ref{fig:speed_Basilisk_contour_iface}c, only then do we see that the velocity magnitude starts to decrease, eventually approaching zero at the contact line, since this is a steady-state case. This is also seen in Figure \ref{fig:speed_python_contour_iface}, where we plot isolines of the velocity magnitude: in the zoomed-out image \ref{fig:speed_python_contour_iface}(a), the contour values increase as the contact line is approached, whereas in the zoomed-in image they decrease. Hence the accelerating flow observed by \citet{Clarke_black_white} and \citet{Kistler_PhD} is not contradicting the slip model; it reflects the outer inertial structure, while the slip model regularises the flow only inside the slip-length region.

\begin{figure}
\centering
\includegraphics[width=\textwidth]{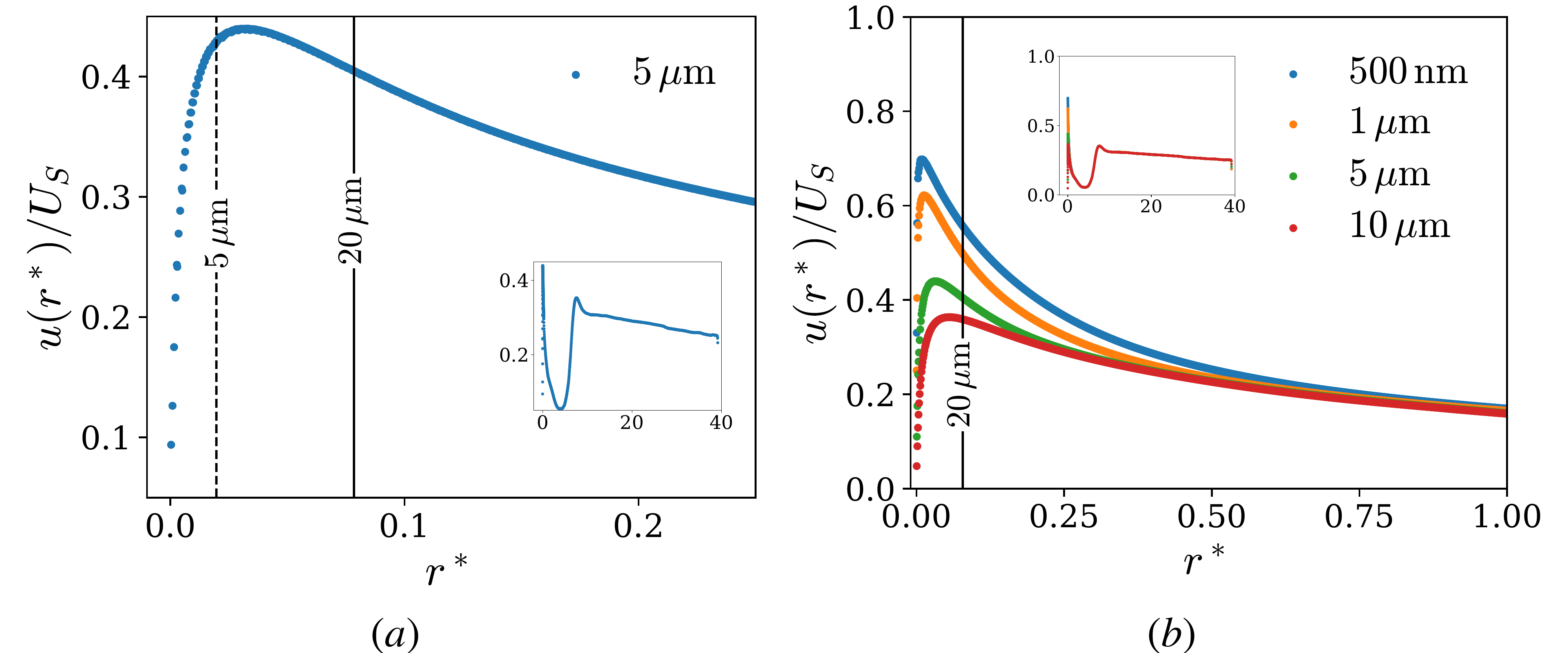}
\caption{The velocity along the interface for $\Reyn = 20$ and $\Ca = 0.7$ for (a) $5\,\mu$m slip length and (b) various slip lengths as indicated in the top-right inset. The inset figures (inset in (a), top-left inset in (b)) represent the full-scale behaviour and the main plot is a zoomed-in version near the contact line. The vertical $20\,\mu$m line is the resolution of the experimental visualisation in \cite{Blake_1999,Clarke_black_white}. $U_{S}$ represents the plate velocity and $r^*$ is the distance scaled with the coated film thickness.}
\label{fig:velocity_iface_FINAL}
\end{figure}

\subsubsection[Velocity along the interface]{The velocity along the interface: The Varma solution\footnote{In this section, the velocity is always defined in the reference frame of the contact line.}}
\label{sec:velocity_along_iface}

We now examine the effect of inertia quantitatively. We plot the velocity along the interface in Figure~\ref{fig:velocity_iface_FINAL}(a). It is seen that (inset) the velocity initially has a slowly increasing plug-flow-like behaviour. It enters the neck region where it increases slightly, then enters the bump region where it slows down and finally starts to increase rapidly. This is in exact coherence with the experimental observations of \citet{Clarke_black_white}. When the distance from the contact line is less than a few microns, below the experimental resolution of \citet{Clarke_black_white}, we see the effect of the slip boundary condition: the velocity starts to decrease and finally goes to zero at the contact line.

The rapid increase before entering the slip region is caused by inertia. The IC-SFW solution of \citet{Varma} for an advancing contact line with a no-slip boundary condition predicts an increase in velocity along the interface reaching the SFW value at the contact line. Since we have a non-zero slip length, we see that the increase in velocity is not up to the SFW limit, but to a certain value below it. As we further reduce the slip length, we see an increase in the maximum velocity value, as shown in Figure~\ref{fig:velocity_iface_FINAL}(b).

\begin{figure}
\centering
\includegraphics[width=0.9\linewidth]{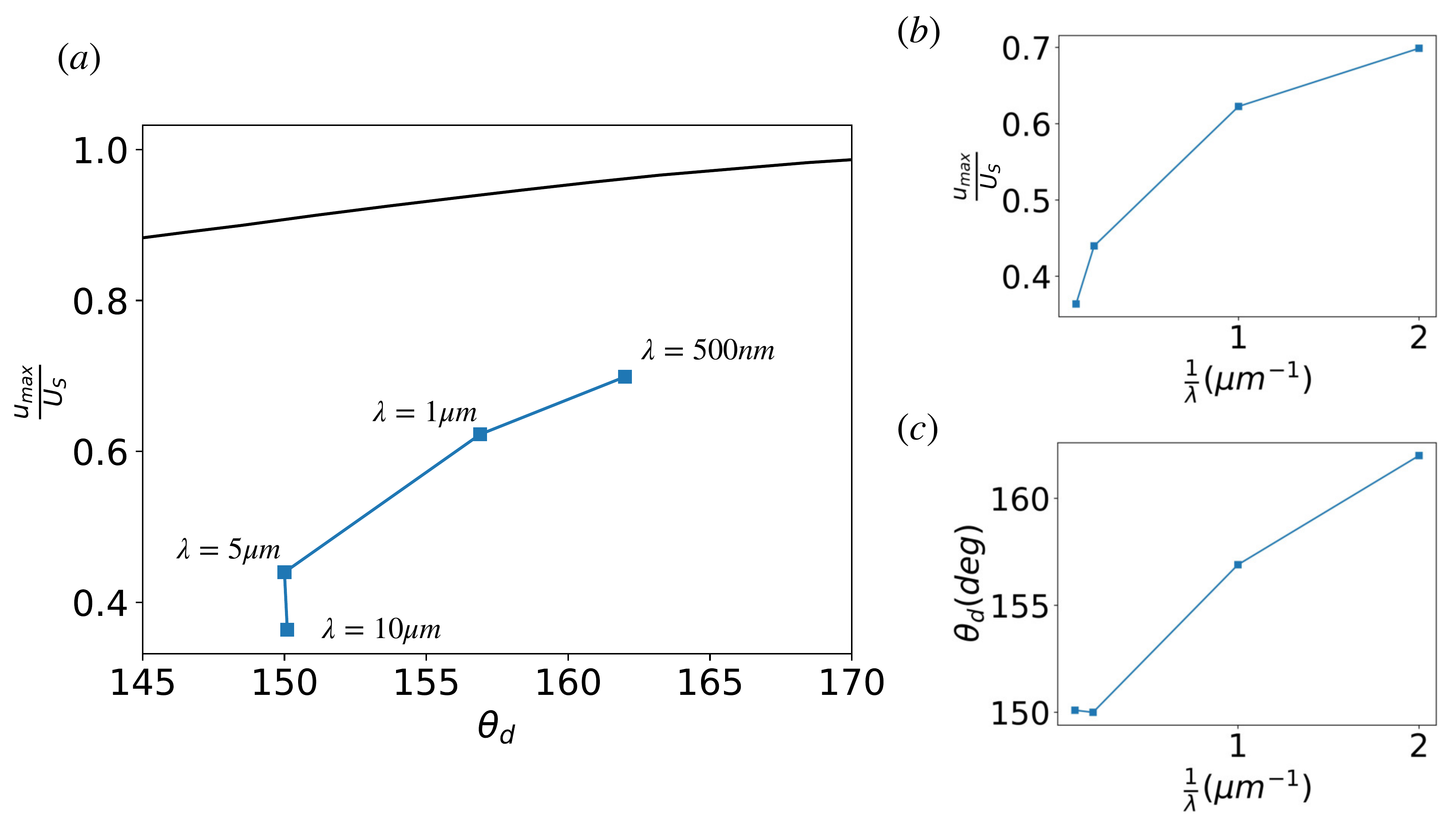}
\caption{(a) The maximum value of the velocity along the interface $u_{max}$ as a function of dynamic contact angle $\theta_d$ measured at the inflection point for various slip lengths $\lambda$. The solid black line represents the theoretical value calculated from the SFW solution. In this SFW solution we set the wedge angle equal to $\theta_d$. The SFW solution serves as an upper bound for $u_{max}$. Values are scaled by the solid plate speed $U_{S}$. Plots (b) and (c) represent the variation of $u_{max}$ and $\theta_d$ with the slip length respectively. The local Reynolds number based on the inflection-point distance is $\bar{r}_{inf}=2.36$ for $\lambda = 10\,\mu$m and $\bar{r}_{inf}=0.44$ for $\lambda =500\,\mathrm{nm}$.}
\label{fig:Data_plot_Vel_iFace_Stokes_curtain}
\end{figure}

In Figure \ref{fig:Data_plot_Vel_iFace_Stokes_curtain} we plot the maximum value of the velocity as a function of slip length and also the dynamic contact angle measured at the inflection point. Figure \ref{fig:Data_plot_Vel_iFace_Stokes_curtain}(b) indicates that the maximum velocity increases as the slip length decreases and Figure \ref{fig:Data_plot_Vel_iFace_Stokes_curtain}(c) shows that the dynamic contact angle measured at the inflection point also increases as the slip length decreases. In the SFW solution with no-slip boundary condition, the constant value of the velocity along the interface increases with wedge angle. Taking the wedge angle equal to the inflection-point angle of our simulation, the constant value of the velocity along the interface calculated from the SFW solution should serve as an upper bound for the maximum value obtained in our simulations. In Figure \ref{fig:Data_plot_Vel_iFace_Stokes_curtain}(a) we compare the numerical maximum velocity with the analytically obtained SFW value. We see that the maximum is indeed bounded by the SFW value, and the difference decreases as the slip length is reduced.

\begin{figure}
\centering
\includegraphics[width=\linewidth]{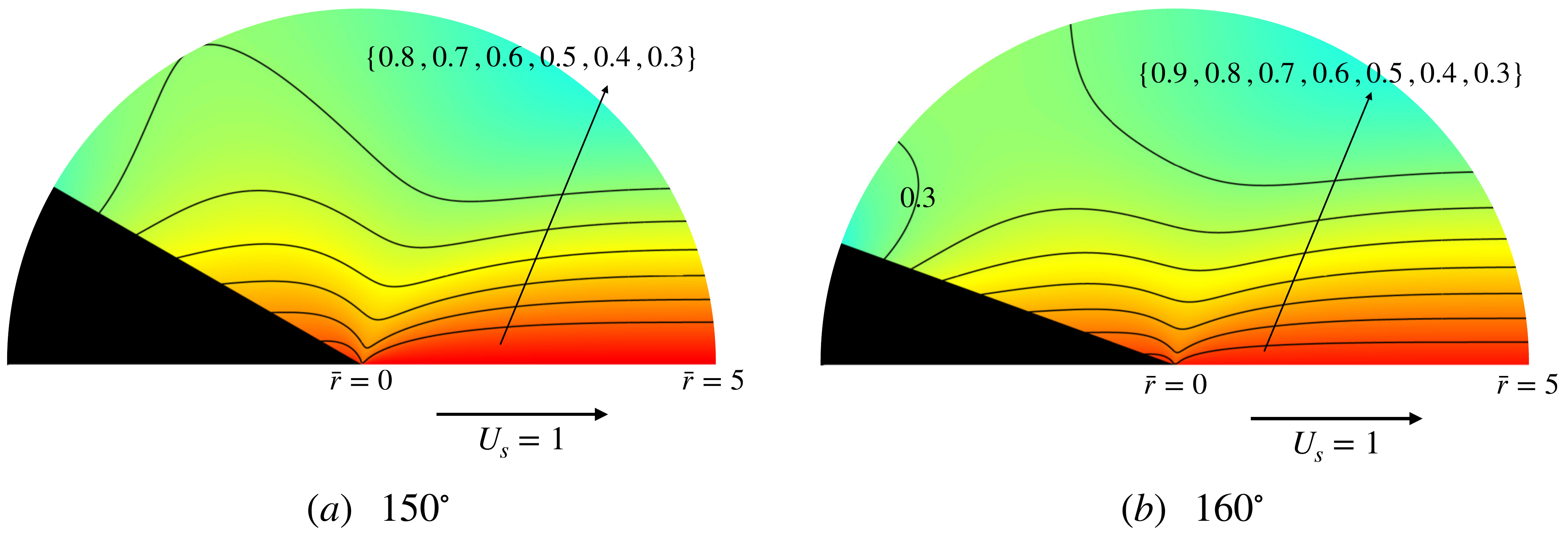}
\caption{The numerical solution for IC-SFW for two different wedge angles. The contour map shows the coloring by the velocity magnitude and the isolines show that the flow accelerates as we approach the contact line. The bottom wall is the no-slip wall moving with velocity $U_{S}=1$ and the inclined edge represents the free surface. Simulations are 2D and $\Bar{r}$ is the local Reynolds number used as dimensionless distance.}
\label{fig:Basilisk_wedge_corner}
\end{figure}

We now show that at an apparent scale outside the slip region (a few tens of microns), the flow can be described by the IC-SFW, with the wedge angle equal to the inflection-point angle. The steady-state Navier--Stokes equation can be written in streamfunction formulation as
\begin{equation}
\Delta^2 \Psi (\Bar{r} , \theta) = \dfrac{1}{\Bar{r}} \left( \frac{\partial \Psi}{\partial \theta} \frac{\partial (\Delta \Psi)}{\partial \Bar{r}} - \frac{\partial \Psi}{\partial \Bar{r}} \frac{\partial (\Delta \Psi)}{\partial \theta} \right),
\label{eq:Varma_inertia_Stokes}
\end{equation}
where $\Bar{r}$ is the local Reynolds number as well as the dimensionless distance and $\theta$ is the angular coordinate. The contact line is at $\Bar{r}=0$. Note that equation \eqref{eq:Varma_inertia_Stokes} with RHS$=0$ is the SFW equation. \revised{A local asymptotic solution of this equation was obtained} by \citet{Varma} in a wedge with $\theta = 0$ (interface) to $\theta = \alpha$ (no-slip moving solid plate). Once the streamfunction is obtained for $\Bar{r}$ and $\theta$, the velocity along the interface can be calculated as a function of $\Bar{r}$. \revised{They resolved the singularity at $\alpha=0.715\pi\,(\simeq129^\circ)$ analytically, but the analytical predictions were restricted to angles at or below this value because, at larger angles, the leading correction depends on disturbance amplitudes set by the external flow.}

Hence we numerically solve the IC-SFW equation \eqref{eq:Varma_inertia_Stokes} using \textit{Basilisk}. We use the embedded-boundary method in 2D to create a wedge and solve the full Navier--Stokes equations until a steady state is reached. The no-slip boundary condition is imposed using a Dirichlet boundary condition on the domain boundary (solid wall). The inclined embedded boundary is used to represent the locally flat fluid--fluid interface rather than a second solid wall; we therefore impose a symmetry (free-slip) condition there, corresponding to no flow through the interface and zero normal gradient of the tangential velocity. Details of the implementation of Dirichlet and Neumann boundary conditions on embedded boundaries can be found in \url{https://basilisk.fr/src/embed.h}; see also \citet{Ghigo_Popinet_Wachs_2021}. These boundary conditions are based on the scheme of \citet{Johansen1998ACG}. The numerical IC-SFW solutions for wedge angles $150^\circ$ and $160^\circ$ are shown in Figure \ref{fig:Basilisk_wedge_corner}. We see that the flow field accelerates up to the contact line. The final value of the velocity along the interface obtained at the contact line for the IC-SFW solution must match the SFW solution. This is verified in Appendix \ref{appex:SFW_IC-SFW}.

We extract the velocity along the interface from the numerical IC-SFW solution for the wedge angle equal to the inflection-point angle and compare it against the full curtain setup. Since we have a slip boundary condition on the entire solid surface, it is not reasonable to use the actual solid speed values directly in this comparison. IC-SFW is a no-slip solution, whereas due to slip at the fluid-solid surface the effective velocity felt by the fluid in the full curtain is smaller than the solid plate velocity. Hence, we scale the velocity field with the $u_{max}$ obtained from Figure \ref{fig:Data_plot_Vel_iFace_Stokes_curtain}a (blue squares). Note that in the physical setup the slip length would be a few nanometers, so the effect of slip would be negligible at a few tens of microns; in our setup the slip length is of the order of microns, so this effective velocity correction must be taken into account.

\begin{figure}
\centering
\includegraphics[width=\linewidth]{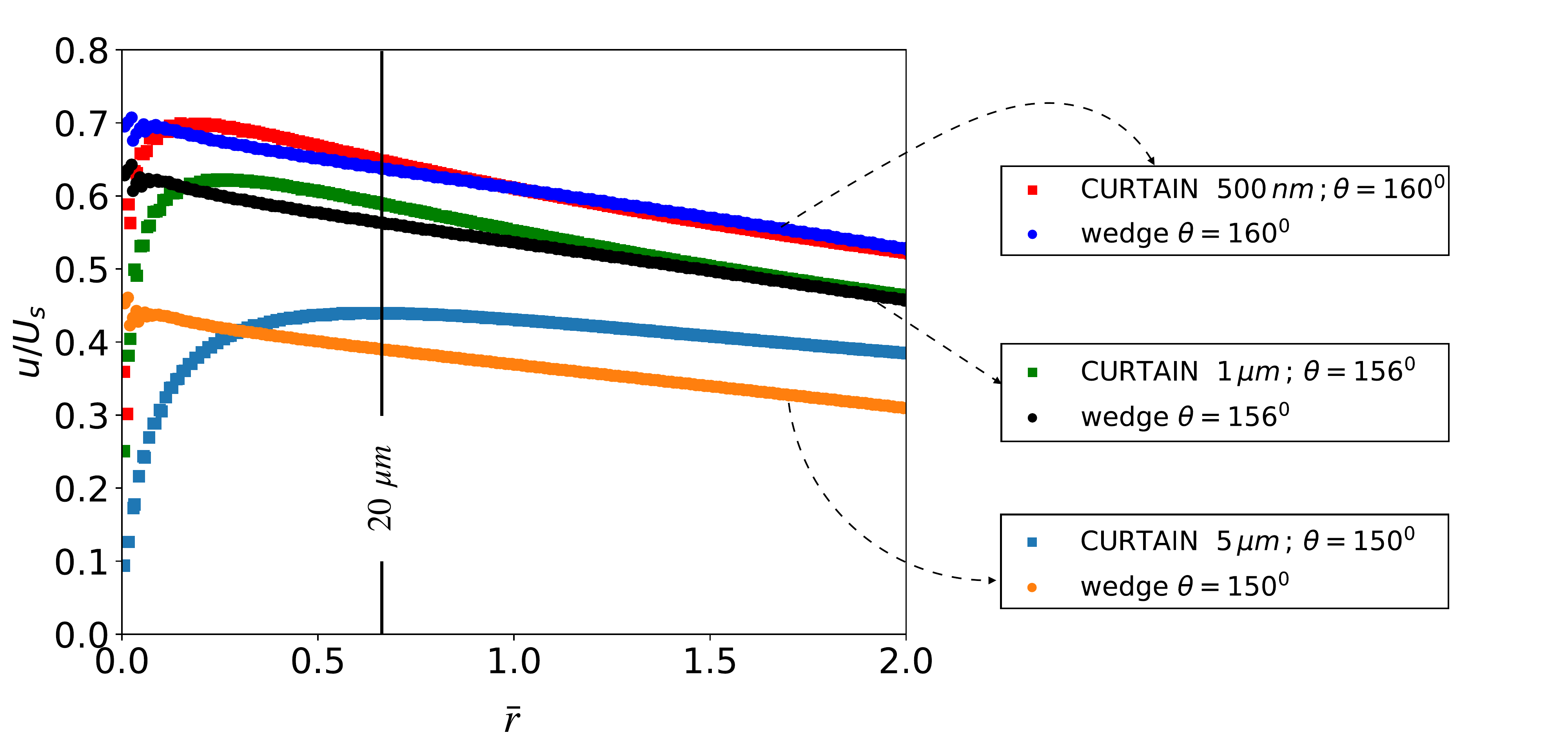}
\caption{Comparison for the velocity along the interface for the numerically obtained IC-SFW wedge solution against the full curtain-coating simulation. The value of $\theta$ is the inflection-point angle for the full curtain-coating setup and is also the corresponding wedge angle for the IC-SFW solution.}
\label{fig:Curtain_wedge_basilisk}
\end{figure}

The final comparison for the scaled velocity of the IC-SFW numerical solution with the full curtain solution is shown in Figure \ref{fig:Curtain_wedge_basilisk}. We see that as the slip length is decreased, the full curtain velocity profile approaches the numerical IC-SFW solution at the micrometre scale. We cannot perform resolved-slip simulations at slip lengths smaller than $500\,\mathrm{nm}$ due to the time-step limitation.

\begingroup

\subsubsection{The outer interface: a sessile-drop-like shape}
\label{subsec:outer_shape}

\begin{figure}
 \centering
 \includegraphics[width=0.99\textwidth]{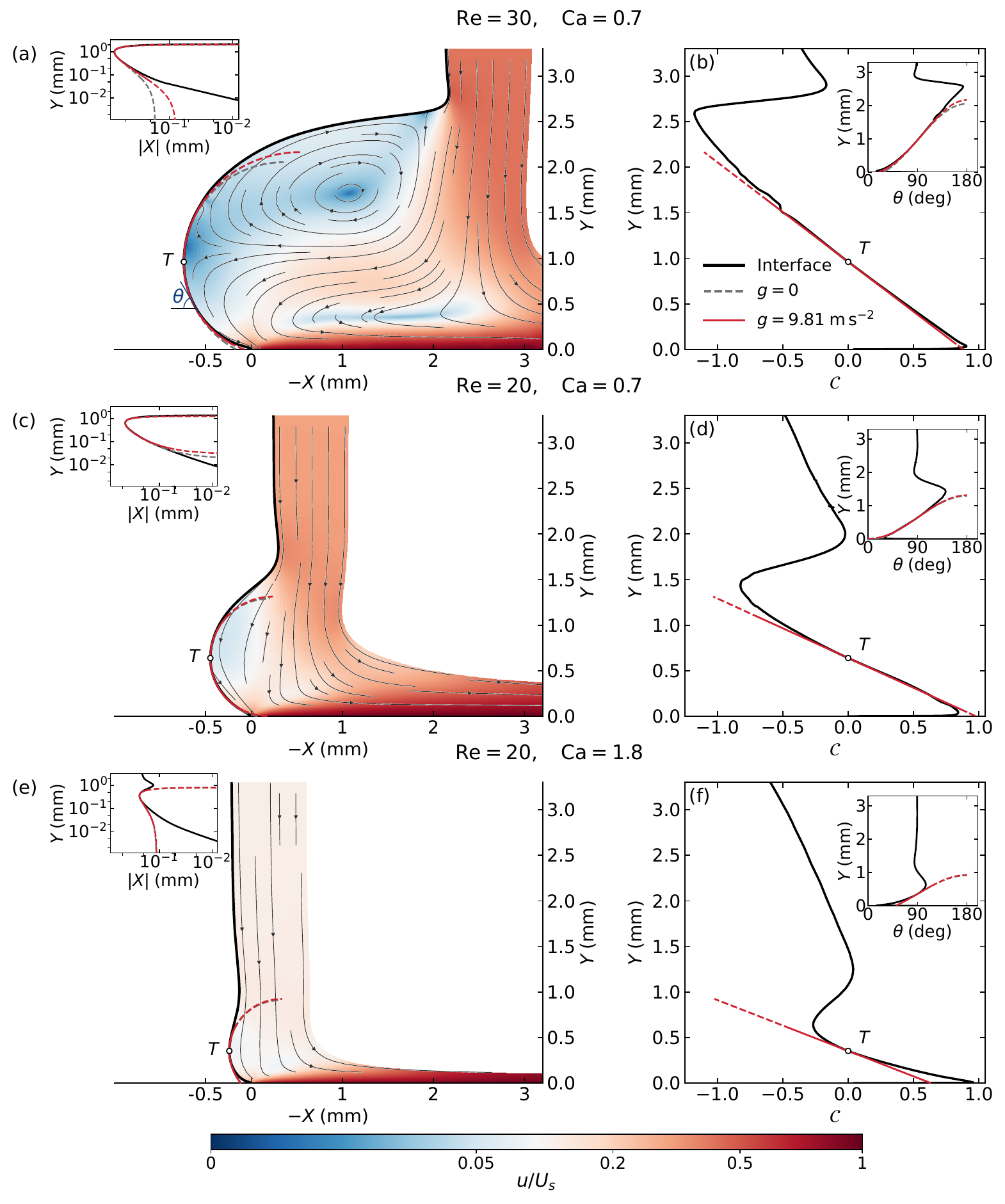}
 \caption{
 Outer shape and flow for $(\Reyn,\Ca)=(30,0.7)$, $(20,0.7)$ and $(20,1.8)$,
 from top to bottom; $\lambda=10\,\mu\mathrm{m}$.
 (a,c,e) Simulated interface (black), streamlines and speed $u/U_{S}$,
 with $u=|\boldsymbol{u}|$ and $U_{S}$ the substrate speed of each case.
 Identical physical axes and equal aspect ratios show the shrinking heel;
 the horizontal coordinate is displayed as $-X$ to retain rightward
 substrate motion. Insets show the positive-$X$ branch on logarithmic axes.
 Red and grey curves are the gravity-dependent Young--Laplace arc and its
 circular ($g=0$) limit, respectively, constructed from the same measured
 position, tangent and curvature at $T$. Solid red segments denote the
 comparison regions and dashed segments extend the same solution.
 (b,d,f) Measured $\mathcal{C}$--$Y$ relation and prediction
 \eqref{eq:outer_YL_C}, with tangent-angle comparisons inset.
 Details of the construction and positional errors are given in
 appendix~\ref{app:outer_shape_methods}.}
 \label{fig:outer_YL_shape}
\end{figure}

The outer heel provides a possible connection between the accelerating
IC-SFW region and the large-scale curtain. In
Figure~\ref{fig:outer_YL_shape}, we compare three cases with different heel sizes, at the same slip
length, $\lambda=10\,\mu\mathrm{m}$. We use the same physical axes and equal aspect ratios in
Figures~\ref{fig:outer_YL_shape}(a,c,e) for a direct visual comparison
of the heels. Since heel formation is favoured at higher $\Reyn$ and
lower $\Ca$, we choose $(\Reyn,\Ca)=(30,0.7)$ as a developed-heel case.
The heel then becomes smaller when $\Reyn$ is reduced to $20$ at the
same $\Ca$, and shrinks further when $\Ca$ is increased to $1.8$.
The common colour bar shows the speed relative to the substrate speed of
each case, $u/U_{S}$.

We first look at the developed heel for $(\Reyn,\Ca)=(30,0.7)$ in
Figure~\ref{fig:outer_YL_shape}(a). We see a broad region of low speed
inside the heel together with a clear recirculating flow. The faster
through-flow in the falling curtain and the coated film passes around
this slowly moving liquid. Near the free surface, the tangential velocity
reverses in the low-speed region above the geometric turn $T$, giving a
surface stagnation region. This suggests that the slowly moving part of
the heel may have a near-equilibrium interface shape. We therefore test
whether a local sessile-drop solution describes this region.

Let $Y$ increase upward and $X$ towards the heel, and let $\theta$ be the
tangent angle measured from the horizontal, as shown in
Figure~\ref{fig:outer_YL_shape}(a). We orient the arclength $s$ from the
contact line towards the curtain, so that the signed planar curvature is
$\kappa=d\theta/ds$. We define the turn $T=(X_T,Y_T)$ as the point where
the interface is vertical, $\theta_T=\pi/2$, and denote the curvature
there by $\kappa_T$. We use $T$ simply as a convenient reference point
at the liquid interface of the heel.

For a hydrostatic capillary interface \citep[\S1.9]{Batchelor1967}, the
Young--Laplace relation gives
\begin{equation}
 \kappa(Y)=\kappa_T-G(Y-Y_T),\qquad
 G=\frac{(\rho_l-\rho_g)g}{\sigma}.
 \label{eq:outer_YL_kappa}
\end{equation}
Thus gravity makes the curvature vary linearly with height, while
$\kappa_T$ contains the reference-pressure contribution. We take the
position and curvature at $T$ directly from the simulated interface.
Once this reference point is selected, there are no fitting
parameters left to fit over the comparison region. For $g=0$,
equation~\eqref{eq:outer_YL_kappa} reduces to a circular arc of radius
$1/\kappa_T$. We show both constructions in
Figures~\ref{fig:outer_YL_shape}(a,c,e); gravity is present in all the
simulations.

To test the sessile drop solution better, we derive an integrated version of ~\eqref{eq:outer_YL_kappa}. Integrating
equation~\eqref{eq:outer_YL_kappa} gives
\begin{equation}
 \mathcal{C}(Y)
 \equiv \cos\theta-\frac{G}{2}(Y-Y_T)^2
 =-\kappa_T(Y-Y_T).
 \label{eq:outer_YL_C}
\end{equation}
Thus, if the sessile-drop description holds, the measured
$\mathcal{C}$--$Y$ curve should follow a straight line with
$d\mathcal{C}/dY=-\kappa_T$. We already know this slope from the measured
curvature at $T$, so we do not fit the straight line to the
$\mathcal{C}$ data. We compare the prediction with the simulation in
Figures~\ref{fig:outer_YL_shape}(b,d,f), and the insets show the same
comparison directly in terms of the tangent angle.

For the largest heel, $(\Reyn,\Ca)=(30,0.7)$, we see good agreement over
a wide region around $T$ where the interface on $\mathcal{C}$--$Y$ coordinate is linear. The red arc follows the simulated
interface in Figure~\ref{fig:outer_YL_shape}(a), while the measured
$\mathcal{C}$--$Y$ curve follows the predicted straight line in
Figure~\ref{fig:outer_YL_shape}(b). Inset shows that the $g=0$ circular arc is already
quite close to the gravity-dependent solution, but a small difference is
visible, particularly for the $\Reyn=30$ case. This is consistent with the
large vertical extent of the heel, since the hydrostatic change in
curvature across a height $\Delta Y$ scales as $G\Delta Y$.

When we reduce $\Reyn$ from $30$ to $20$ at $\Ca=0.7$, the heel
contracts, as seen by comparing Figures~\ref{fig:outer_YL_shape}(a) and
\ref{fig:outer_YL_shape}(c). We still find a region around $T$ that is
well described by the sessile-drop-like arc, as also shown by the
$\mathcal{C}$ comparison in Figure~\ref{fig:outer_YL_shape}(d).
However, the gravity-dependent and $g=0$ curves are now almost on top of
each other, including in the inset comparisons. The smaller vertical
extent of the heel means that gravity changes the curvature much less
across this region.

Finally, increasing $\Ca$ to $1.8$ at $\Reyn=20$ suppresses the heel even
further, as shown in Figure~\ref{fig:outer_YL_shape}(e). The
gravity-dependent and circular arcs are again almost indistinguishable,
consistent with the much smaller vertical extent of the heel. However,
the agreement with the simulated interface is weaker and the region over
which the $\mathcal{C}$ relation is approximately linear is smaller, as
seen in Figure~\ref{fig:outer_YL_shape}(f).

We therefore see a clear progression in
Figure~\ref{fig:outer_YL_shape}. For the developed heel, a substantial
region around $T$ is well described by the sessile-drop-like solution.
As the heel becomes smaller, this region also becomes smaller, and the agreement gets weaker. At the same
time, the difference between the gravity-dependent and circular solutions
decreases because the vertical extent of the heel is reduced.

The sessile-drop-like region therefore provides a natural outer
description of the heel beyond the accelerating IC-SFW region discussed
in the previous section. Together, the two results
suggest a flow structure in which the accelerating IC-SFW region connects
to a slowly moving, sessile-drop-like outer heel.

\par
\endgroup

\subsection{Hydrodynamic assist and stability in the curtain configuration}
\label{subsec:compare_exp_Liu}

\begin{figure}
\centering
\begin{subfigure}[b]{0.49\textwidth}
  \centerline{\includegraphics[width=\textwidth]{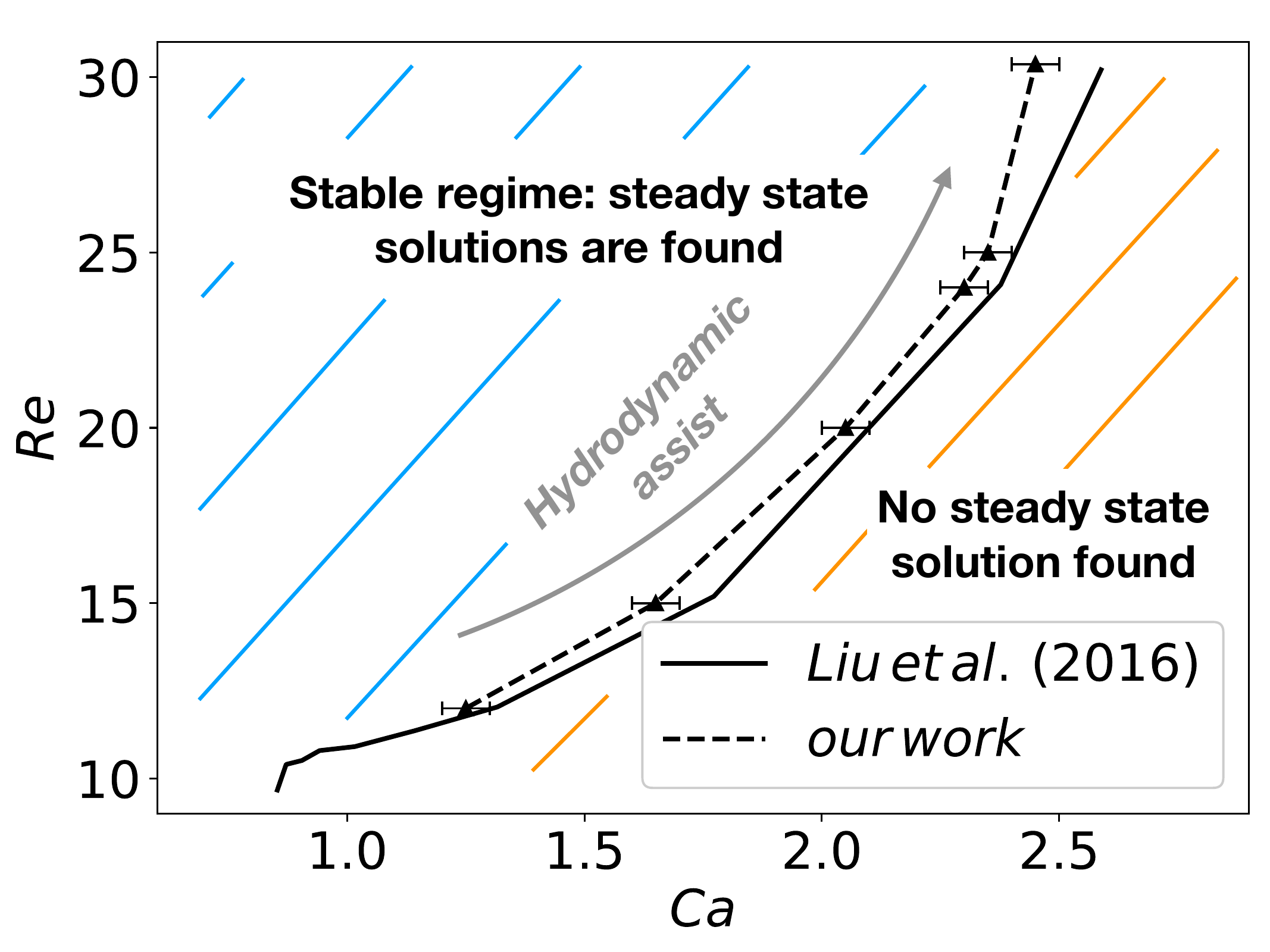}}
  \caption{}
\label{fig:stability_window_Liu}
\end{subfigure}
\hfill
\begin{subfigure}[b]{0.49\textwidth}
\centering
\includegraphics[width=\textwidth]{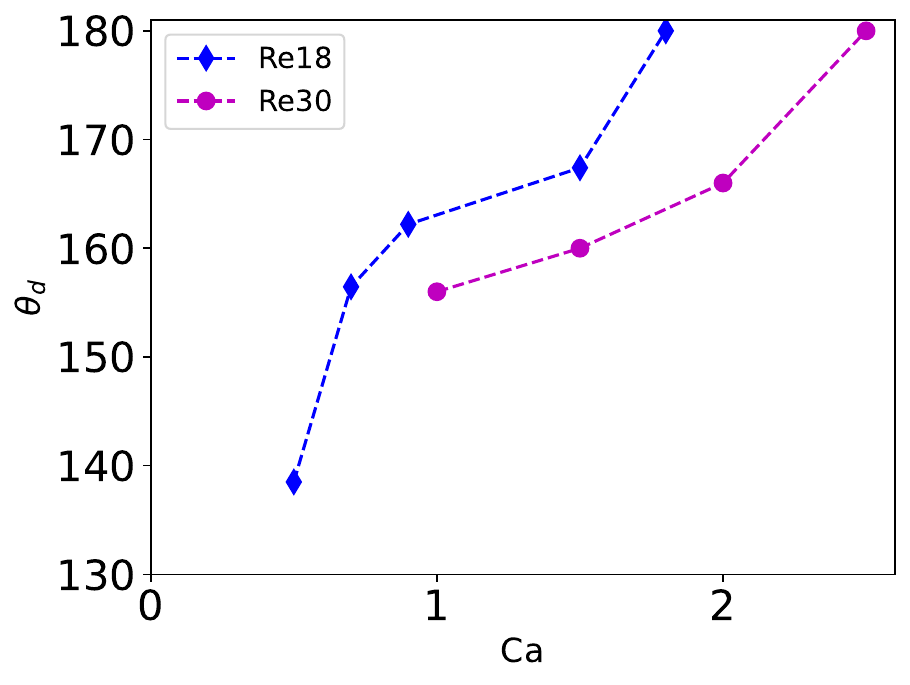}
\caption{}
\label{fig:macro_micro_reduced}
\end{subfigure}
\caption{(a) Comparison of the stability window obtained in the current simulations against the previous simulations of \citet{Liu16}. The curves indicate the critical capillary number above which wetting failure takes place and no steady-state solutions are found. The left portion of the bifurcation curve refers to the stable regime where steady-state solutions are found. (b) Macroscopic contact angle $\theta_d$ measured at the inflection point as a function of $\Ca$ for different $\Reyn$ for the reduced model. The values on the y-axis are in degrees and a variation of more than $20^\circ$ is clearly seen for various $\Reyn$.}
\end{figure}

We now present the results of our simulations using the reference parameters of \citet{Liu16}. We have the following parameters (symbols are as per Figure \ref{curtain_Tomas} and $\sigma$ is the surface tension):
\begin{equation*}
\begin{aligned}
\rho_l &= \SI{1000}{\kilogram\per\meter^{3}},
&\quad \mu_l &= \SI{25}{\milli\pascal\second},
&\quad d_c &= \SI{1e-3}{\meter}, \\
\rho_g &= \SI{1.2}{\kilogram\per\meter^{3}},
&\quad \mu_g &= \SI{0.018}{\milli\pascal\second},
&\quad h_c &= \SI{1e-2}{\meter}, \\
\sigma &= \SI{70}{\milli\newton\per\meter},
&\quad \theta_m &= \SI{90}{\degree},
&\quad \lambda &= \SI{1e-5}{\meter}.
\end{aligned}
\end{equation*}

These parameters are useful for two reasons. First, the relatively large slip length $\lambda$ makes the computations cheaper and, more importantly, enables \emph{resolved-slip} simulations in which the slip region is explicitly resolved rather than represented implicitly by a grid-dependent effective slip length. This allows us to probe the near-contact-line structure and to perform controlled convergence studies. The limitation is that such a large $\lambda$ is not intended for quantitative comparison with experiments; in particular, at high $\Ca$ it does not guarantee that inertia is negligible at the slip-length scale, unlike the physical nanometric-slip situation. For this reason we later perform slip-length reduction down to our computational limits.

Figure~\ref{fig:stability_window_Liu} compares our stability boundary with that reported by \citet{Liu16}. We recover the hydrodynamic-assist trend: increasing the feed flow (larger $\Reyn$) delays wetting failure, i.e.\ increases the critical capillary number $\Ca_{cr}$ over a substantial portion of parameter space. The agreement is notable because \citet{Liu16} modelled the gas stresses using a reduced one-dimensional description, whereas we solve the full two-phase Navier--Stokes equations; the close match in the low-$\Reyn$, low-$\Ca$ regime supports the accuracy of their gas-stress model in that limit.

We next consider the apparent contact angle measured at the interface inflection point, located a few tens of micrometres ($20$--$50\,\mu$m) from the contact line. At this position the curvature changes sign and the local slope varies slowly, making it a natural proxy for the experimentally observed macroscopic angle \citep{Voinov}. In slip-based descriptions with a prescribed microscopic angle, the inflection point represents an intermediate scale
\[
\lambda < r_{inf} < r_{\text{macro}},
\]
where capillary pressure reverses sign to deflect the gas phase away from the contact line and sustain steady wetting. In experiments such as \citet{Blake_1999}, where $\lambda$ is nanometric, this location lies well below optical resolution ($\sim 20\,\mu$m). In the present reduced model ($\lambda=10\,\mu$m), however, the inflection-point angle provides a convenient macroscopic proxy that can be consistently extracted from the simulations and compared with experimentally reported apparent angles.

Figure~\ref{fig:macro_micro_reduced} shows that this angle depends not only on the capillary number but also on the large-scale Reynolds number. At fixed $\Ca$, variations of several tens of degrees are observed, consistent with the trends reported by \citet{Blake_1999}. The crossing of constant-$\Reyn$ curves indicates a non-monotonic dependence on $\Reyn$. The dashed black curve marks the stability boundary: wetting failure does not necessarily occur when the angle reaches $180^\circ$. Instead, for a given $\Reyn$, the angle increases with $\Ca$ up to a critical value $(\le 180^\circ)$ beyond which no steady solution exists. The observation of $\theta_{inf}<180^\circ$ at the transition is consistent with both molecular-dynamics and continuum studies \citep{Keeler_Blake_thin_Ca,Liu16}.

Having characterised the steady solutions and their stability boundary, we now use the reduced model to verify that the resolved-slip numerics capture the known inner logarithmic singular structure, and to assess how the stability window changes when the slip length is varied.

\subsubsection{The curvature singularity}

The Navier-slip formulation is known theoretically to possess a logarithmic (integrable) pressure and curvature singularity at the moving contact line \citep{Hocking_YK,Olivier_SZ}. The objective of this subsection is to verify that the present resolved-slip simulations correctly capture this inner asymptotic structure.

\begin{figure}
\centering
\begin{subfigure}[b]{0.49\textwidth}
\centering
\includegraphics[width=\textwidth]{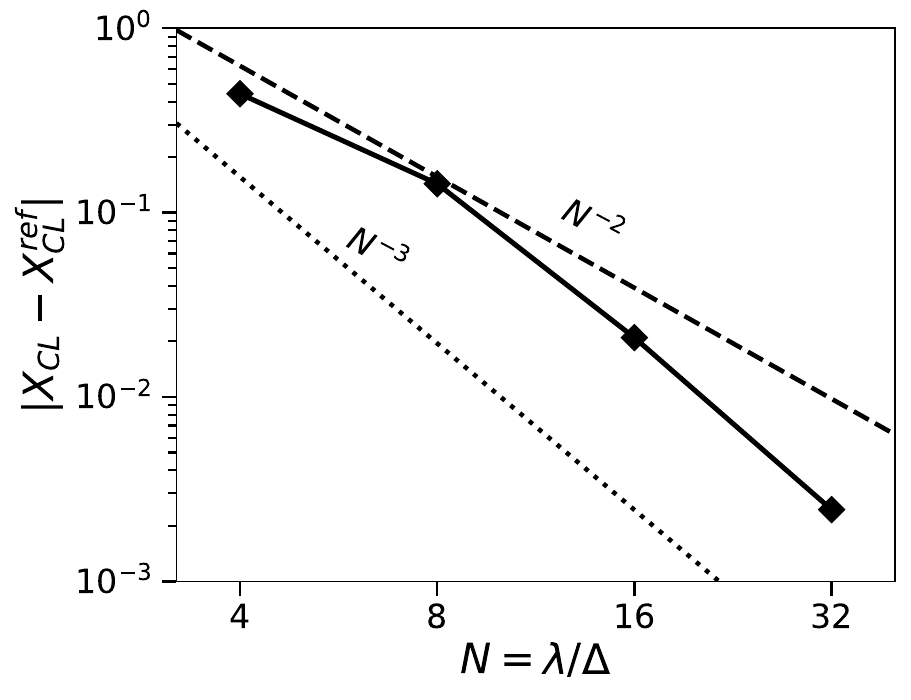}
\caption{}
\end{subfigure}
\hfill
\begin{subfigure}[b]{0.49\textwidth}
\centering
\includegraphics[width=\textwidth]{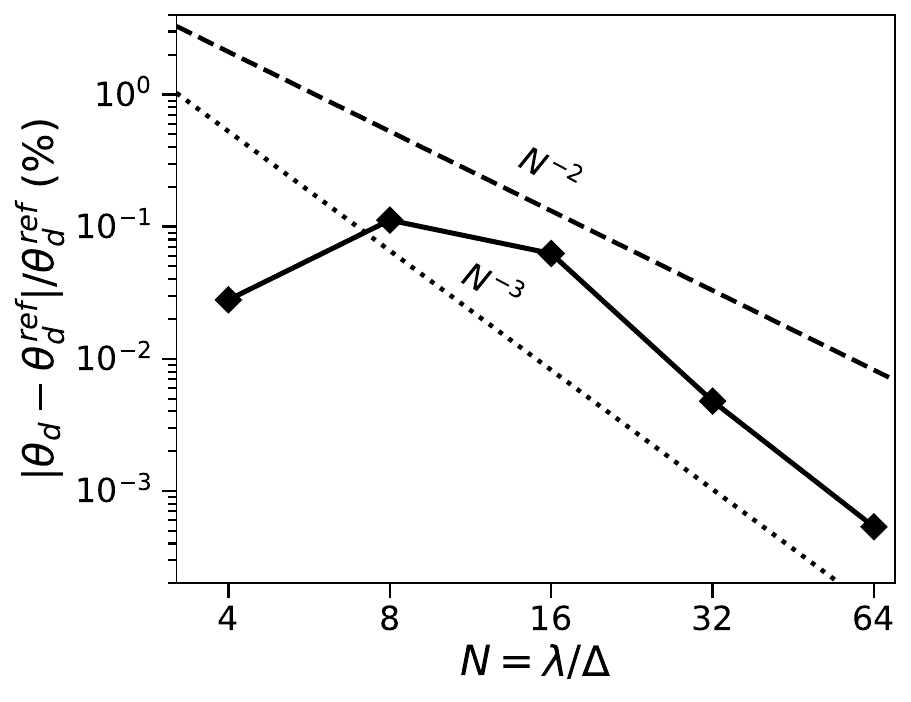}
\caption{}
\end{subfigure}
\caption{Convergence of the resolved-slip simulations for the reduced curtain model at $\Reyn=20$ and $\Ca=0.7$, for which a steady-state solution is obtained.
(a) Contact-line position error, $|X_{CL}-X_{CL}^{ref}|$, as a function of the number of grid points per slip length, $N=\lambda/\Delta$. The reference contact-line position $X_{CL}^{ref}$ is taken from the simulation with $N=64$.
(b) Relative error in the apparent contact angle measured at the inflection point, $100|\theta_d-\theta_d^{ref}|/\theta_d^{ref}$, where $\theta_d^{ref}$ is taken from the simulation with $N=128$.
Dashed lines indicate the reference slopes $N^{-2}$ and $N^{-3}$.}
\label{fig:convergence_study}
\end{figure}

Because the slip length is explicitly resolved, the solution converges with the number of grid points per slip length $N=\lambda/\Delta$, unlike approaches where the slip is implicitly determined by the grid size ($\sim \Delta/2$). Figure~\ref{fig:convergence_study} shows convergence of both the contact-line position and the inflection-point angle with approximately third-order accuracy, demonstrating that the near-contact-line region is numerically resolved. Additional refinement-level diagnostics, including the progressive improvement of the VoF reconstruction and the local angle/curvature behaviour near the contact line, are given in Appendix~\ref{app:resolved_slip_diagnostics}.

\begin{figure}
\centering
\includegraphics[width=\textwidth]{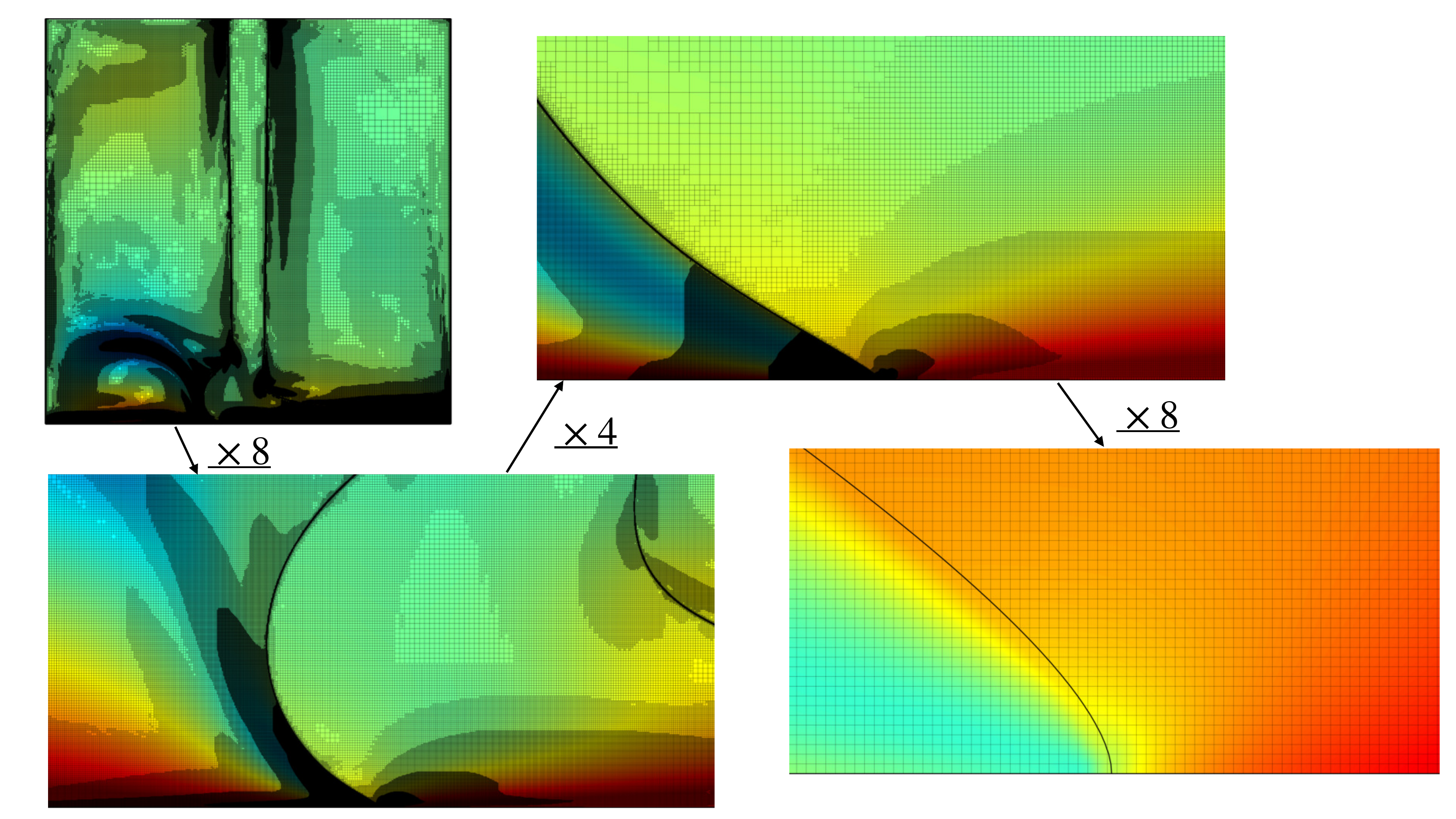}
\caption{Successive zoom-in near the contact line in a steady-state simulation of $\Reyn = 20$ and $\Ca = 0.7$. Mesh refinement can be seen and the background is colored by the horizontal velocity.}
\label{fig:zoom-in_Re20_Ca0.7}
\end{figure}

As the grid is refined, the contact angle at the wall converges to the imposed value of $90^\circ$, while its gradient grows without bound, as shown in Appendix~\ref{app:resolved_slip_diagnostics}. Thus the interface remains locally straight at the contact line while becoming non-smooth there. This is precisely the behaviour expected from the integrable logarithmic singularity of the Navier-slip model in the Stokes-flow limit. A successive zoom-in near the contact line is shown in Figure~\ref{fig:zoom-in_Re20_Ca0.7}.

\begin{figure}
\centering
\begin{subfigure}[b]{0.49\textwidth}
\centering
\includegraphics[width=\textwidth]{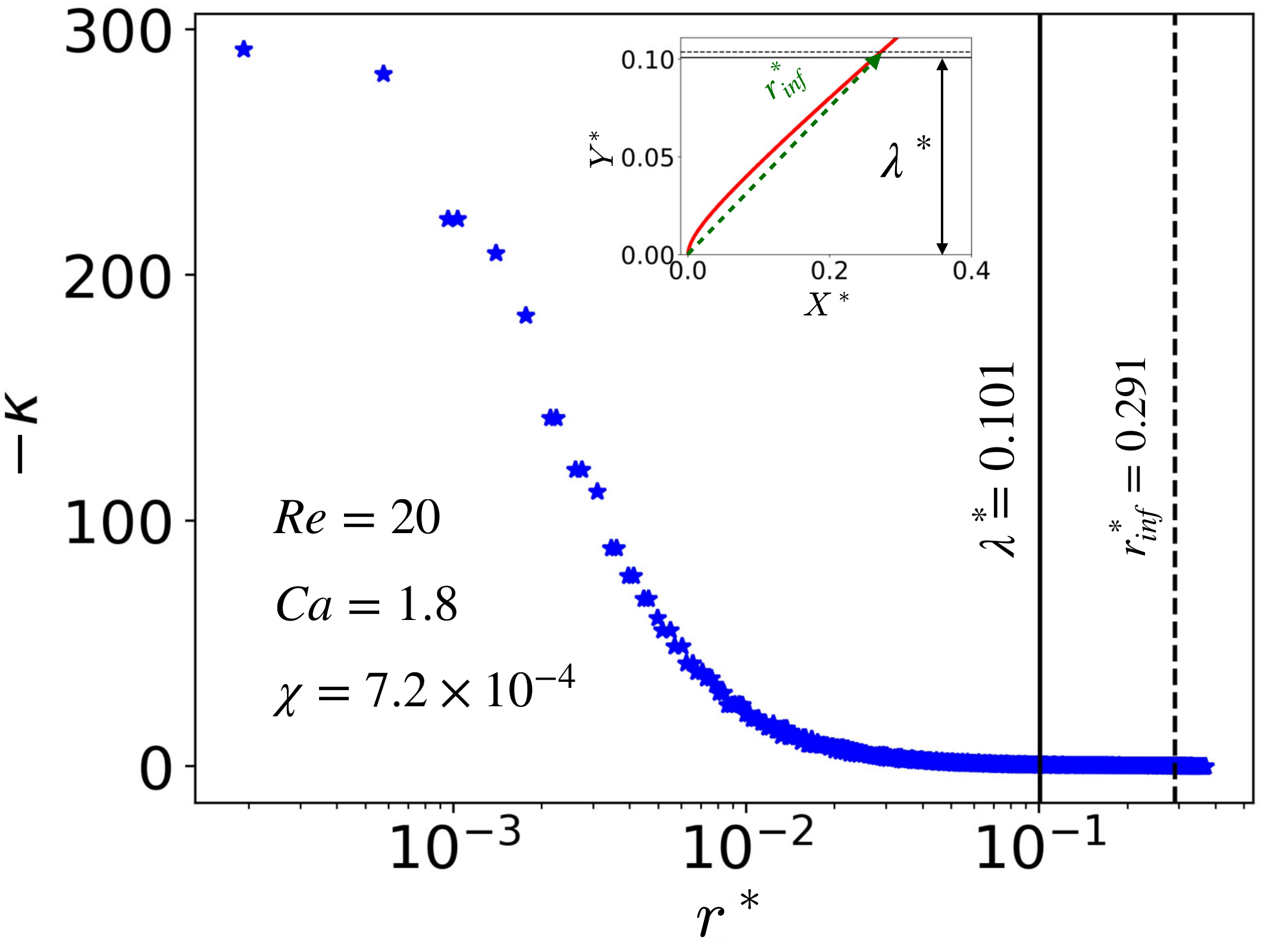}
\caption{}
\label{fig:log_kappa_Re20Ca1.8}
\end{subfigure}
\hfill
\begin{subfigure}[b]{0.49\textwidth}
\centering
\includegraphics[width=\textwidth]{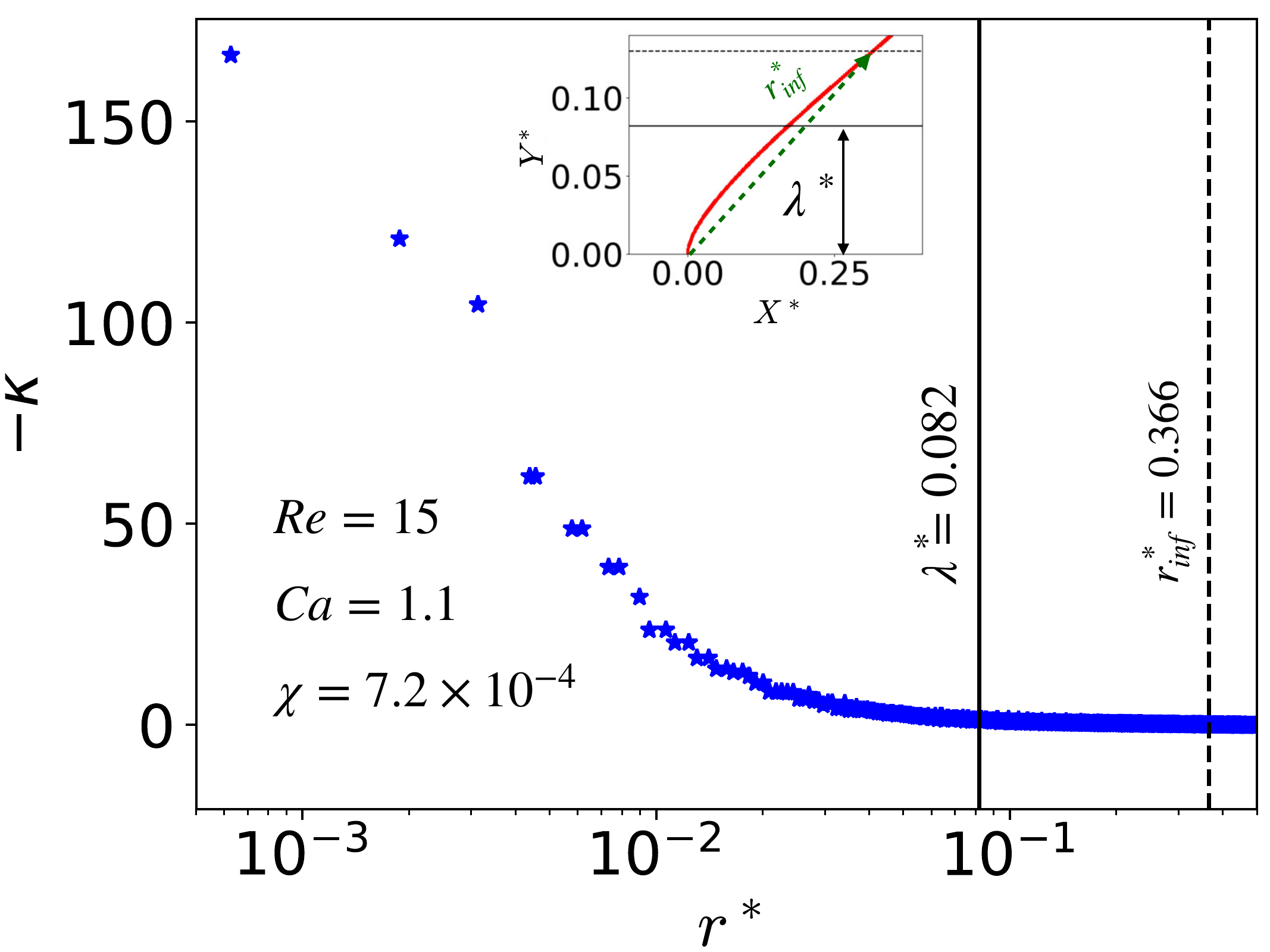}
\caption{}
\label{fig:log_kappa_Re15Ca1.1}
\end{subfigure}
\caption{Logarithmic divergence of the curvature at the contact line. The inset shows the zoomed-in plot of the interface in the vicinity of the contact line $(0,0)$. The starred quantities $X^*$, $Y^*$, $r^*$, $\lambda^*$ and $r_{inf}^*$ denote lengths scaled by the coated film thickness, with $r^*$ measured from the contact line. The dashed line represents the inflection-point distance $r_{inf}^*$ and the solid line represents the slip length $\lambda^*$. The physical slip length is $\lambda=10\,\mu\mathrm{m}$ in both panels, and $\chi$ denotes the viscosity ratio. The plots are for (a) critical capillary number and (b) sub-critical capillary number.}
\label{fig:log_kappa_divergence}
\end{figure}

Figure~\ref{fig:log_kappa_divergence} directly shows the logarithmic divergence of curvature. Because the singularity is integrable, the wall shear stress remains finite and steady solutions exist despite the divergence. Notably, near the stability limit the height of the inflection point approaches the slip length, indicating a possible loss of scale separation at the onset of wetting failure.

\subsubsection{Film entrainment during wetting failure}

\begin{figure}
\centering
\begin{subfigure}[b]{0.49\textwidth}
\centering
\includegraphics[width=\textwidth]{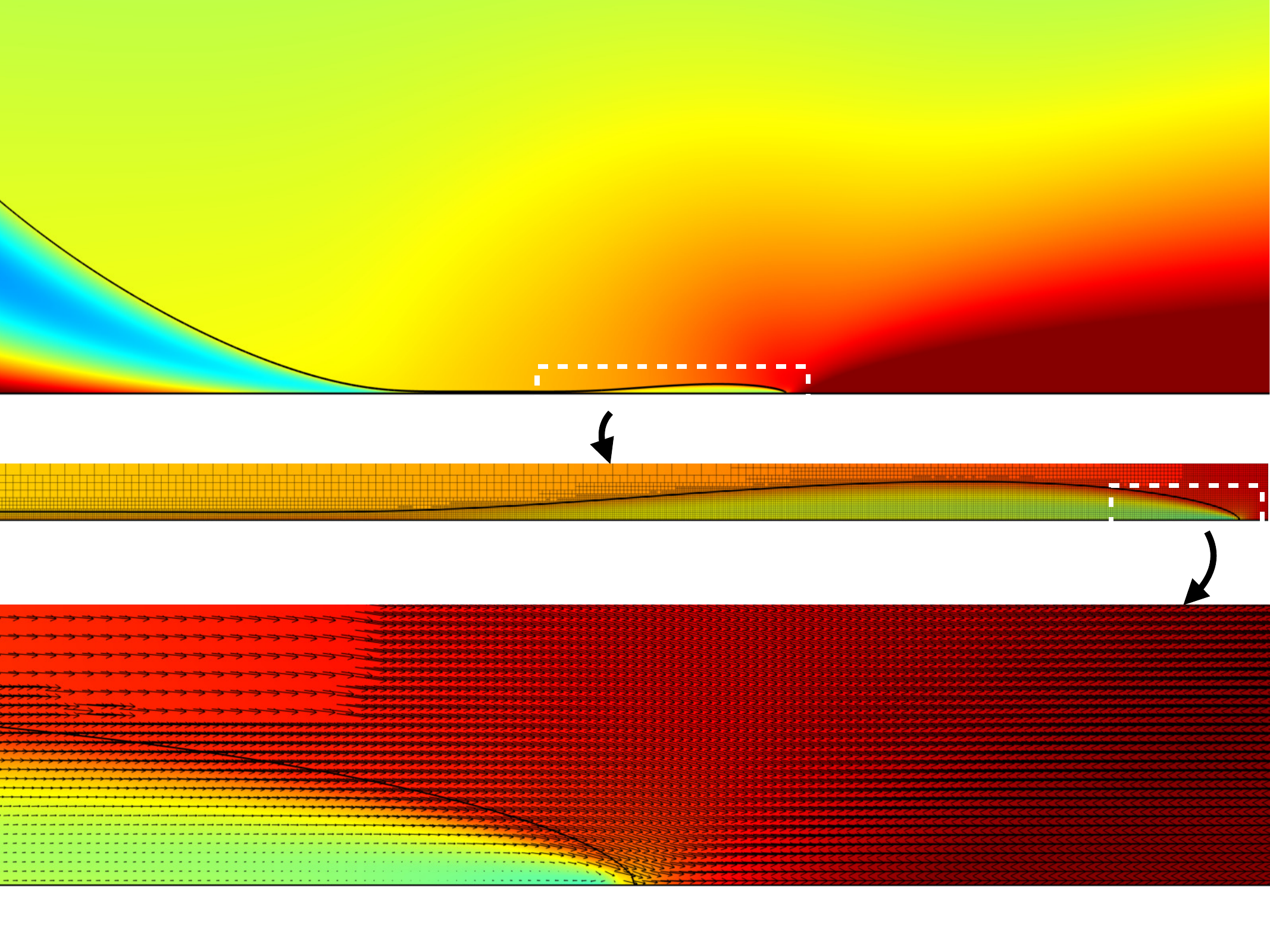}
\caption{}
\label{fig:wetting_failure_sim_entrainment}
\end{subfigure}
\hfill
\begin{subfigure}[b]{0.49\textwidth}
\centering
\includegraphics[width=\textwidth]{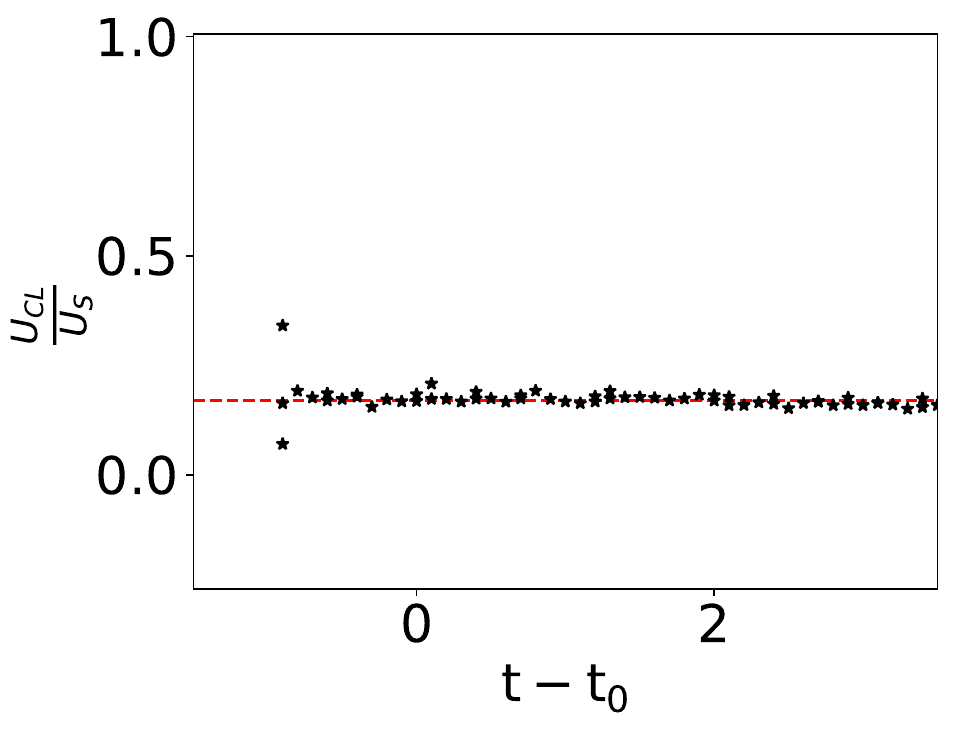}
\caption{}
\label{fig:wetting_failure_sim_vel}
\end{subfigure}
\caption{Wetting failure case for $\Reyn = 20$ and $\Ca = 2.0$ with $10\,\mu$m slip length and 64 grid points per slip length. (a) Zoom in near the advancing air-film cusp. The background is colored by $u$ velocity and velocity vectors are shown in the most zoomed-in version. (b) The contact-line velocity (cusp tip) as a function of shifted time, $t-t_0$. Time is non-dimensionalised by $h_{\mathrm{inf}}/U_{S}$, where $h_{\mathrm{inf}}$ is the coated film thickness and $U_{S}$ is the solid-substrate speed; $t_0$ denotes the onset of resolved air-film entrainment. The horizontal red-dashed line is $y = 0.17$.}
\label{fig:wetting_failure_sim}
\end{figure}

We consider an unstable case of the reduced model ($\Reyn=20$, $\Ca=2$). Instead of approaching a steady state, the interface develops an entrained air film with a cusp-like shape that propagates along the plate (Figure~\ref{fig:wetting_failure_sim}). The apparent angle measured at the inflection point becomes reflex, indicating loss of a steady coating solution. A notable feature is that the cusp tip rapidly reaches an approximately constant velocity in the laboratory frame, about $17\%$ of the plate speed (Figure~\ref{fig:wetting_failure_sim_vel}). Wetting failure therefore corresponds to the replacement of a stationary contact line by a steadily propagating entrainment front.

\subsubsection{Effect of the slip length reduction}
\label{subsec:slip_length_reduction_resolved}

Experiments report a non-monotonic stability window, whereas the reduced model with a large slip length yields a monotonic behaviour for hydrodynamic assist on an $\Reyn$--$\Ca$ plane. We therefore progressively reduce the slip length toward physically realistic values to examine how the stability window evolves.

\begin{figure}
\centering
\begin{subfigure}[b]{0.49\textwidth}
\centering
\includegraphics[width=\textwidth]{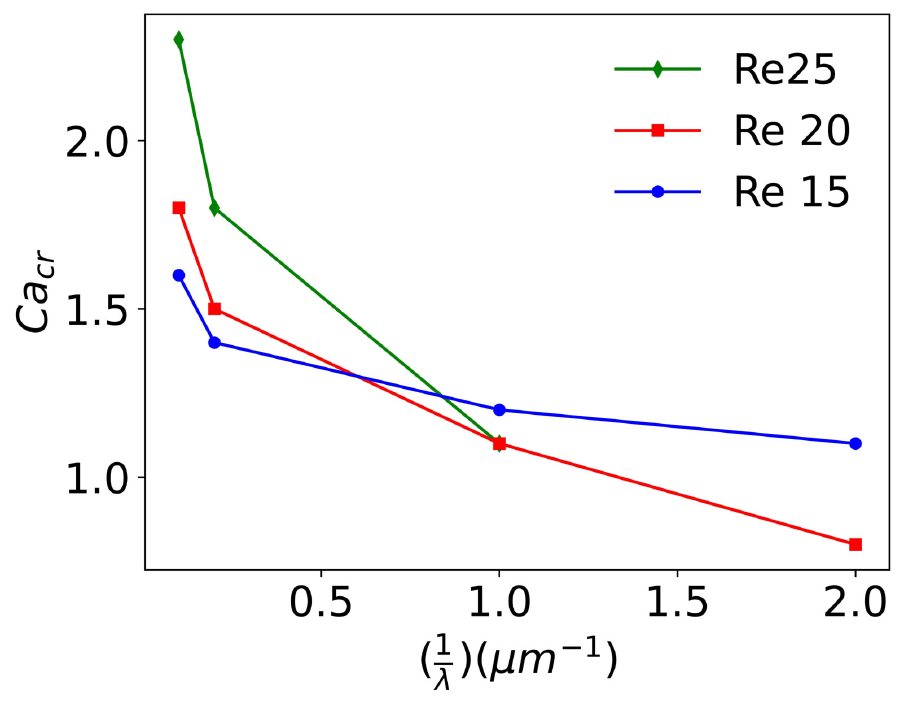}
\caption{}
\label{fig:low_slip_reduced}
\end{subfigure}
\hfill
\begin{subfigure}[b]{0.49\textwidth}
\centering
\includegraphics[width=\textwidth]{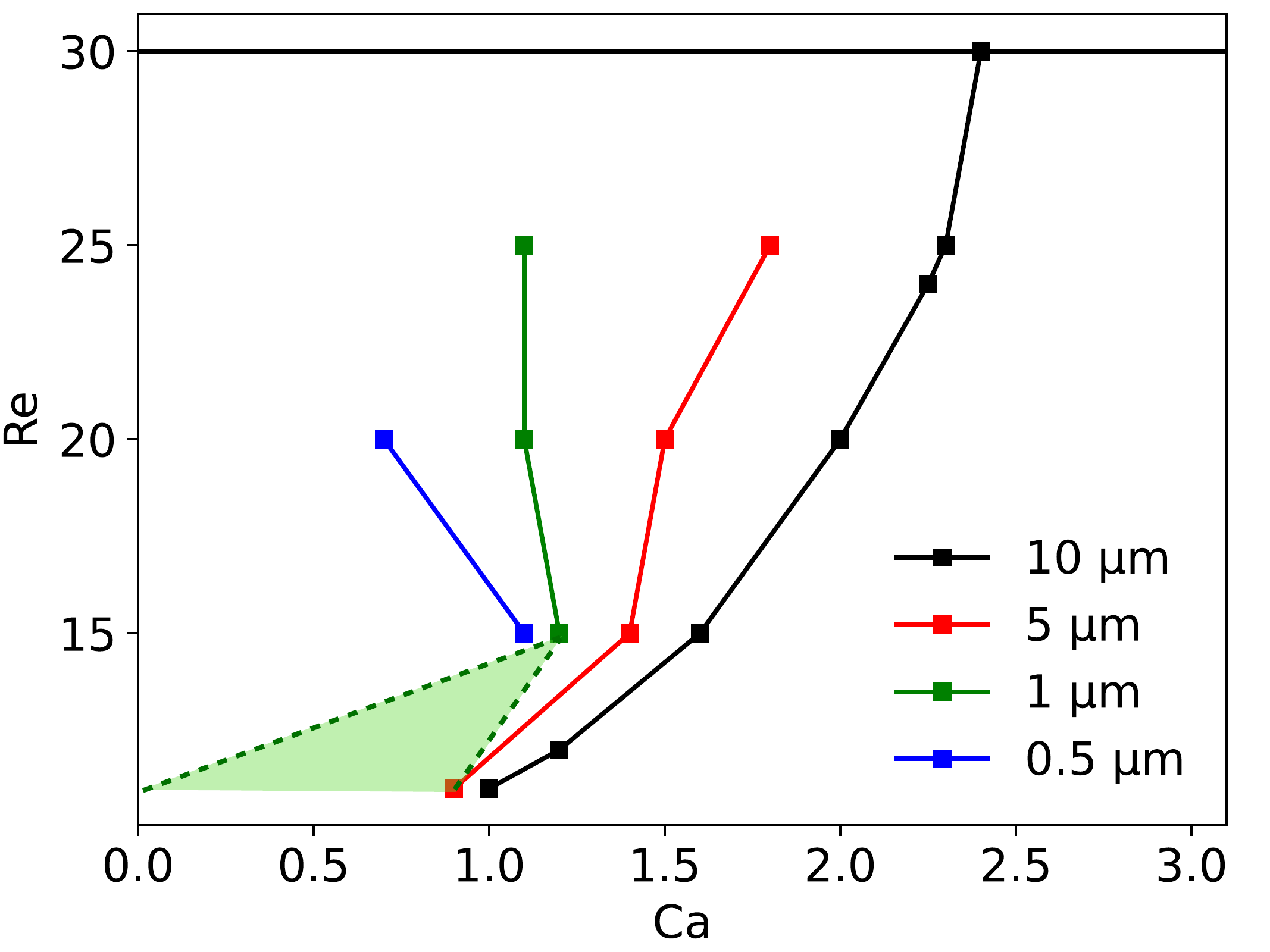}
\caption{}
\label{fig:slip_vary_stability}
\end{subfigure}
\caption{(a) Critical capillary number as a function of slip length for various Reynolds numbers indicated in the legend. (b) Stability window as the slip length is reduced. The shaded green region indicates the possible domain of critical capillary numbers at low $\Reyn$ and low $\Ca$ for the $\lambda=1\,\mu$m case. Direct simulations in this range were not performed because low-$\Reyn$, low-$\Ca$ cases tend to be in the bead-pulling configuration and require an elongated computational domain. For $\lambda=5\,\mu$m and $\Reyn=10$, we obtain $\Ca_{cr}=0.9$; the corresponding critical capillary number for $\lambda=1\,\mu$m must therefore be smaller, giving the shaded region. The figure shows that reducing the slip length makes the stability window non-monotonic.}
\end{figure}

Figure~\ref{fig:low_slip_reduced} shows the evolution of the critical capillary number $\Ca_{cr}$ as the slip length decreases for several Reynolds numbers. As $\lambda$ is reduced, the solution branches cross and the stability boundary becomes non-monotonic: increasing $\Reyn$ does not always increase $\Ca_{cr}$. This behaviour is clearly visible in the stability diagram (Figure~\ref{fig:slip_vary_stability}). The monotonic trend observed for $\lambda=10\,\mu$m disappears at smaller slip lengths ($1\,\mu$m and $500\,$nm), where a non-monotonic stability window emerges.

A complementary study of the influence of the imposed microscopic contact angle on the large-scale angle measured in the heel configuration is reported in Appendix~\ref{app:micro_macro}.

\section{Discussion}
\label{sec:discussion}
The results above separate two levels of interpretation. At the resolved numerical level, the Navier-slip simulations reproduce the reduced-model stability trends, show the expected logarithmic curvature singularity, and recover a non-monotonic stability window as the slip length is reduced. At the physical level, they clarify why the accelerating interfacial velocity observed at micrometric distances from the contact line does not directly contradict slip regularisation at nanometric scales: the experimentally accessible region and the microscopic slip region belong to different parts of the multiscale contact-line structure.

A critical aspect of this study is explaining the accelerating flow observed at scales of a few tens of micrometres from the contact line. In the Stokes-flow limit, a slip-regularised solution predicts that the velocity decreases as the contact line is approached and eventually vanishes at the contact line. This appears, at first sight, to contradict the experimental observation of an increasing interfacial velocity down to optical resolution. The key point is that, in curtain coating at $\Ca=O(1)$, the flow at these observable distances is not in the Stokes-slip inner region. The local Reynolds number, $\Bar{r}=\rho_l |U_{S}| r/\mu_l$, is already $\mathcal{O}(1)$ for $r\sim 10\,\mu$m when $\Ca=1$, so the flow at tens of microns is governed by inertial corrections. This is why the inertially corrected Stokes-flow wedge solution is the relevant comparison. When the wedge angle is taken equal to the angle observed at the inflection point, the IC-SFW solution captures the micrometric velocity increase reasonably well, and the agreement improves as the slip length is decreased.

\revised{The sessile-like outer shape in \S\ref{subsec:outer_shape} suggests how
the developed heel may complement the IC-SFW description. The interfacial
speed falls within the heel, and an arc constructed from the measured
turning-point data reproduces a finite outer region without fitting an
analytical shape parameter there. The region is broader for the larger
heel and contracts as the bump becomes smaller in the cases examined.

Together, these two descriptions also suggest a possible route to a
sub-grid model for the rapidly advancing contact line in the present setup. A coarse simulation would only need to
resolve the heel sufficiently to determine $T$ and $\kappa_T$. These
quantities fix the outer Young--Laplace arc, whose continuation towards
the substrate provides an extrapolated apparent angle for the unresolved
contact-line region. This angle could then set the wedge angle of the
IC-SFW solution, while the inner slip solution connects the inertial
wedge to the microscopic contact angle. In the present calculations we
instead obtain the wedge angle directly from the resolved inflection
point. Testing such a reduced description is left for future work.
}

The present computations remain limited by the accessible separation
of scales. \revised{At our highest resolution, we have a Basilisk refinement level 17, a $1\,\mathrm{cm}$ domain gives a minimum cell width of $\Delta=1\,\mathrm{cm}/2^{17}\simeq76.3\,\mathrm{nm}$.} Thus the full two-phase simulations span about five orders of magnitude, which is enormous for transient simulations with both phases, but still below the full centimetre-to-nanometre ratio of experiments, corresponding to roughly seven orders of magnitude. Since we wanted to resolve the slip region explicitly, most of the study used slip lengths of order microns, allowing sufficient grid points inside the slip region. The slip-length reduction study should therefore be viewed as an attempt to approach the experimental scale separation, while maintaining enough resolution inside the slip length.

The first consequence is about the interpretation of curtain-coating experiments. The experimentally observed acceleration of the interfacial velocity has often been viewed as being in tension with slip-type models, because a Stokes slip solution decelerates inside the slip region \citep{Yulli_review}. The present results show that this inference is not decisive in the curtain-coating regime. A Navier-slip model can regularise the contact-line motion while preserving, at least qualitatively, the wetting-failure behaviour, the interfacial velocity field, and the interface bending up to the experimental resolution. The reason is that measurements at tens of micrometres do not directly probe the microscopic regularisation region; in the present regime, they probe an inertial micrometric region outside it. This also connects with the classical observation of \citet{Dussan_slip_model}: different descriptions of the inner contact-line region can lead to substantially different local flows while preserving the same meniscus-scale behaviour. Thus, observing a given velocity trend at optical resolution does not uniquely identify the microscopic contact-line law. This does not mean that the Navier-slip model is unfalsifiable in a Popperian sense. Rather, it means that the macroscopic observations currently available in this setup do not, by themselves, constitute a decisive falsification of Navier slip or related mobility-law regularisations; a sharper test would require observables that are sensitive to the transition between the inertial micrometric region and the microscopic regularisation region, or independent information about the microscopic contact-line law.

The model with the slip-length Navier boundary condition yields predictions in qualitative agreement with observation: it predicts wetting failure while preserving the inertial velocity increase and the intermediate-scale interface bending. The logarithmic curvature singularity of the Navier-slip formulation is therefore not fatal for the macroscopic question addressed here. It is weak and integrable, and it does not prevent grid convergence or the existence of steady resolved-slip solutions.

\section{Conclusion}
\label{sec:conclusion}

We have presented two-phase simulations of curtain coating governed by the Navier--Stokes equations with a Navier-slip boundary condition and a prescribed grid-scale contact angle. The computations reproduce the reduced-model stability trends of \citet{Liu16}, including hydrodynamic assist, and provide resolved-slip numerical evidence of the logarithmic curvature singularity associated with the Navier-slip formulation. As the slip length is reduced, the simulations also recover a non-monotonic stability window, qualitatively consistent with experimental observations.

A central objective of this work was to clarify the meaning of the experimentally observed increase of interfacial velocity as the contact line is approached. A purely viscous slip-regularised inner solution predicts deceleration within the slip region, whereas experiments show acceleration down to their resolution limit. Our simulations reconcile these observations. They show that the deceleration associated with slip occurs only inside the microscopic inner region, while the experimentally accessible micrometric region lies outside it and is already influenced by inertia. In this observable region, the interfacial velocity is consistent with the inertially corrected wedge solution of \citet{Varma}, with agreement improving as the slip length is reduced.

\revised{The outer heel admits a complementary sessile-drop-like description. A Young--Laplace arc specified by the measured turning-point position, vertical tangent and locally reconstructed curvature reproduces a finite region of the developed heel, without fitting any analytical arc parameter. The parameters that promote a bigger heel, hence also promote a wider region of sessile-drop-like shape. These results thus suggest a structure in which the accelerating IC-SFW region connects to a slowly moving, sessile-like outer heel.}

The conclusion is therefore not that the Navier-slip model is the final microscopic model of moving contact lines. The model remains weakly singular and physically ill posed at sufficiently small scales, leaving room for other contact-line models that may provide a more regular mathematical description while retaining the correct macroscopic behaviour. At the same time, this limitation does not by itself invalidate the macroscopic use of Navier slip: if the singular region lies below molecular scales, the continuum sharp-interface description has already reached its physical limit, and the singularity can be handled as a mathematical regularisation. The present results show that Navier slip remains compatible with the observed macroscopic behaviour in curtain coating, provided the scale separation between the slip region and the experimentally observable inertial region is recognised. A decisive test of the microscopic contact-line law cannot rely only on interface shapes or interfacial velocities measured at scales where the flow is already controlled by outer hydrodynamics. It would require observables in the transition region between the microscopic regularisation scale and the inertial micrometric region, or independent microscopic information, for example from molecular dynamics or sub-micrometric measurements that are demonstrably sensitive to the chosen contact-line law. The present work is therefore a step toward clarifying the range of validity of continuum contact-line models for rapidly advancing flows, and toward full DNS of the curtain-coating experiment.

\bigskip 

We acknowledge the support of the ERC grant TRUFLOW nr 883849. We are grateful for access to the computational facilities of the French CINES (National computing centre for higher education) and the TGCC granted by GENCI under project number A0092B07760. The project also benefited from the PRACE grant TRUFLOW number 2020225418 for a large number of CPU hours in 2021. We thank the technical and administrative teams of these supercomputer centres and agencies for their kind and efficient help.
We also acknowledge the computational grant on the Swiss computer Piz Daint.\\

We would like to thank Michele Pellegrino, Petter Johansson, Sherwin Bagheri, Berk Hess and Gustav Amberg for the monthly group meetings where several ideas on the contact line flowed freely.

Stephane Zaleski would like to thank Yves Pomeau, Jens Eggers and Yulii Shikhmurzaev for detailed discussions.

Yash Kulkarni would like to thank Mathis Fricke, Dieter Bothe and Jens Eggers for the discussions.

\bigskip
\textbf{Declaration of Interests}: The authors report no conflict of interest.

\bigskip

OpenAI's ChatGPT (GPT-6 Astra) was used to assist with grammatical corrections and minor language editing of the manuscript. All scientific ideas, methodology, simulations, results and conclusions are those of the authors. The authors have reviewed and approved the final manuscript and take full responsibility for its accuracy, originality and scientific content.

\appendix

\section{Temporal and spatial discretisation of the Navier--Stokes solver}
\label{appA}

(This is just an overview. The reader is directed to \cite{Basilisk_Popinet_1} for the detailed procedure.) Our Navier--Stokes solver uses a staggered discretisation of the volume-fraction/density and pressure combined with a time-splitting projection method. The discretised form of the equations appears as below:

\begin{equation}
\begin{aligned}
\rho_{n+\frac{1}{2}}\left(\frac{\vec{u}_{*}-\vec{u}_{n}}{\Delta t}+\vec{u}_{n+\frac{1}{2}} \cdot \vec{\nabla} \vec{u}_{n+\frac{1}{2}}\right)
&=
\vec{\nabla} \cdot\left[\mu_{n+\frac{1}{2}}\left(D_{n}+D_{*}\right)\right]
\\
&\quad+\left(\sigma \kappa \delta_{s} \vec{n}\right)_{n+\frac{1}{2}}
+ \rho_{n+\frac{1}{2}}\vec{g}_{n+\frac{1}{2}}.
\end{aligned}
\label{eq:discrete-momentum}
\end{equation}

\begin{equation}
\frac{c_{n+\frac{1}{2}}-c_{n-\frac{1}{2}}}{\Delta t}
+\vec{\nabla} \cdot\left(c_{n} \vec{u}_{n}\right)=0.
\end{equation}

\begin{equation}
\vec{u}_{n+1}
=
\vec{u}_{\star}
-\frac{\Delta t}{\rho_{n+\frac{1}{2}}} \vec{\nabla} p_{n+\frac{1}{2}}.
\label{eq:velocity_pressure_discrete}
\end{equation}

Note that $\vec{u}^{*}$ is a temporary velocity field, which is found in the first step (equation \ref{eq:discrete-momentum}) by ignoring the pressure-gradient term in the momentum equation \eqref{eq:NS-continuum}. Then the pressure-gradient term is added to get the final velocity field. This two-step procedure is known as the predictor--corrector projection method. To solve for the pressure gradient and ensure a divergence-free velocity field, equation \eqref{eq:velocity_pressure_discrete} is coupled with the Poisson equation

\begin{equation}
\vec{\nabla} \cdot\left[\frac{\Delta t}{\rho_{n+\frac{1}{2}}} \vec{\nabla} p_{n+\frac{1}{2}}\right]
=
\vec{\nabla} \cdot \vec{u}_{\star}.
\end{equation}

Now the discrete form of the momentum equation \eqref{eq:discrete-momentum} can be rearranged as follows:

\begin{equation}
\begin{aligned}
\frac{\rho_{n+\frac{1}{2}}}{\Delta t} \vec{u}_{\star}
-\vec{\nabla} \cdot\left[\mu_{n+\frac{1}{2}} \vec{D}_{\star}\right]
&=
\vec{\nabla} \cdot\left[\mu_{n+\frac{1}{2}} \vec{D}_{n}\right]
+\left(\sigma \kappa \delta_{s} \vec{n}\right)_{n+\frac{1}{2}}
\\
&\quad+\rho_{n+\frac{1}{2}}\vec{g}_{n+\frac{1}{2}}
+\rho_{n+\frac{1}{2}}
\left[
\frac{\vec{u}_{n}}{\Delta t}
-\vec{u}_{n+\frac{1}{2}} \cdot \vec{\nabla} \vec{u}_{n+\frac{1}{2}}
\right].
\end{aligned}
\end{equation}

This scheme is solved by the second-order upwind Bell--Collela--Glaz advection scheme \citep{Bell_Collela}.

\begin{figure}
    \centering
    \includegraphics[width=1\linewidth]{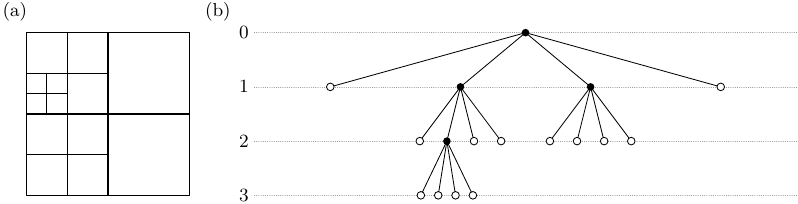}
    \caption{(a) Illustration of quadtree adaptive mesh refinement. (b) The tree representation.}
    \label{fig:spatial_discrete_schemetic}
\end{figure}

Space is discretised using a quadtree partitioning in 2D (octree in 3D), as shown in Figure \ref{fig:spatial_discrete_schemetic}. All variables are collocated at the centre of each square discretisation volume. Consistently with a finite-volume formulation, the variables are interpreted as the volume-averaged values for the corresponding discretisation volume. A projection method \citep{Popinet_Gerris_2003} is used for the spatial discretisation of the pressure-correction equation and the associated divergence in the Poisson equation.

\section{Numerical IC-SFW solution and matching with the inner slip region}
\label{appex:SFW_IC-SFW}

\begin{figure}
    \centering
    \includegraphics[width=\linewidth]{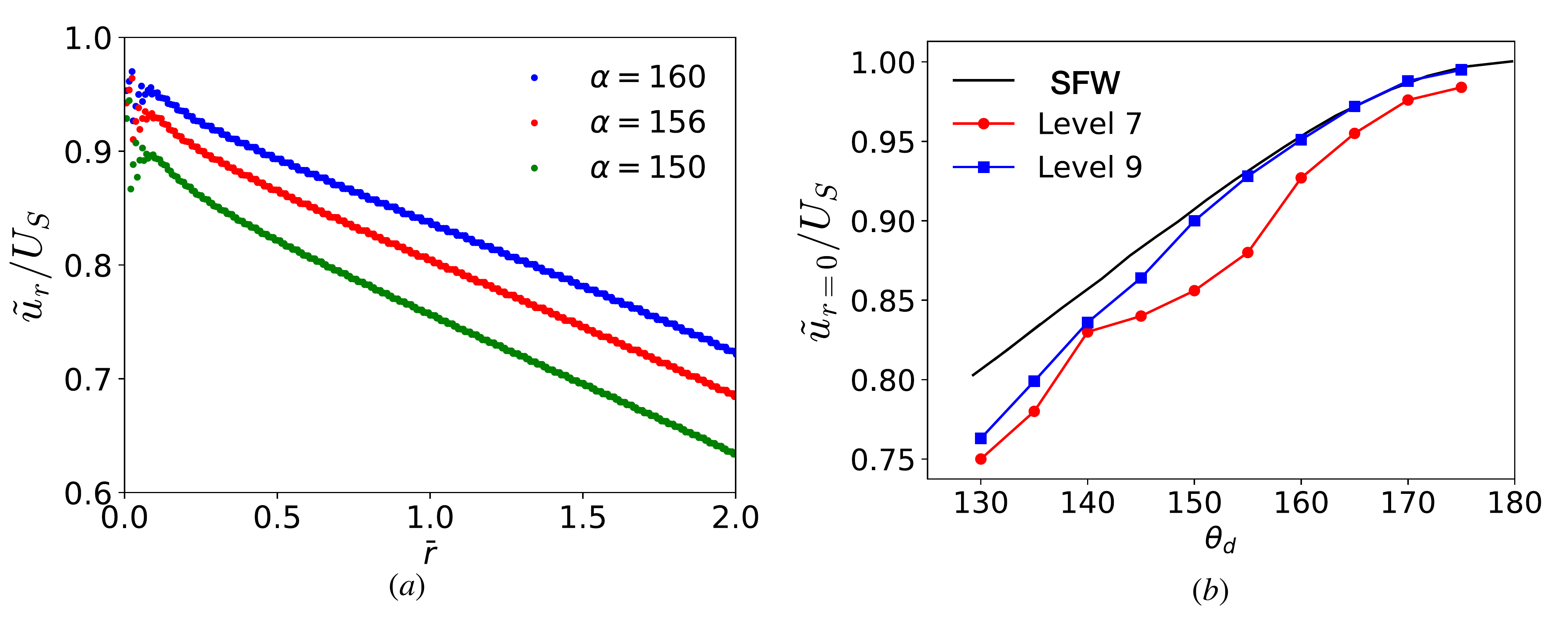}
    \caption{(a) The velocity along the interface for the Basilisk IC-SFW solution for three wedge angles as a function of local Reynolds number $\Bar{r}$. Note that the noise at $\Bar{r} = 0$ is numerical as numerics do not allow a discontinuous velocity field. (b) The value of the velocity at $\Bar{r} = 0$ (five grid cells away from the contact line to avoid noise at the wall) in the IC-SFW solution of (a) is extracted for each wedge angle and compared against the analytical SFW solution. Level 7 represents a mesh of $2^7 \times 2^7 = 16384$ cells in the computational domain up to $\Bar{r}=10$ and Level 9 corresponds to a mesh of $2^9 \times 2^9 = 262144$ grid points.}
    \label{fig:basilisk_IC-SFW_all_angles}
\end{figure}

\begin{figure}
\centering
\begin{subfigure}{\linewidth}
    \centering
    \includegraphics[width=0.7\linewidth]{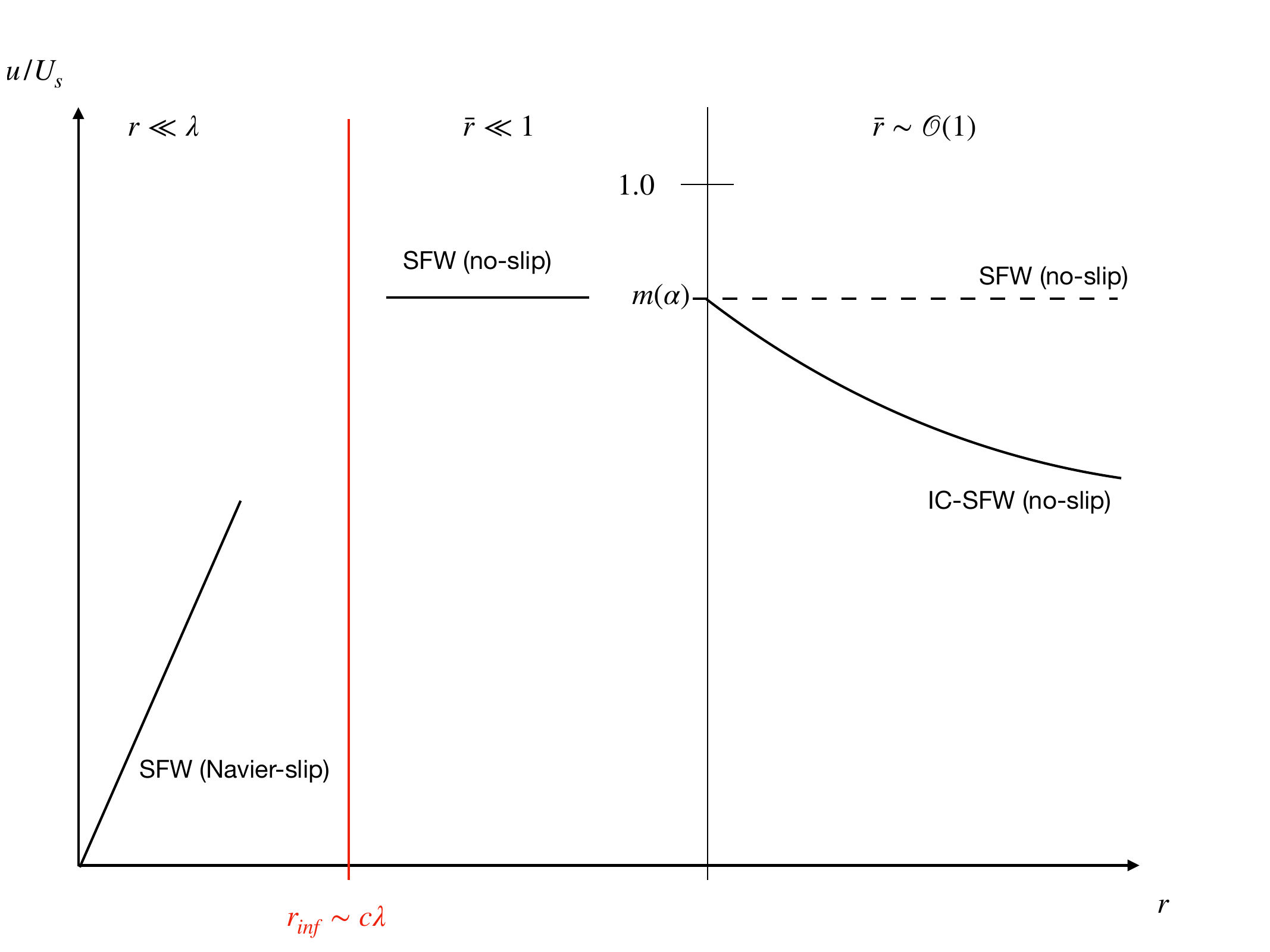}
    \caption{}
    \label{fig:Schematic_IC_SFW_a}
\end{subfigure}

\vspace{1em}

\begin{subfigure}{\linewidth}
    \centering
    \includegraphics[width=0.7\linewidth]{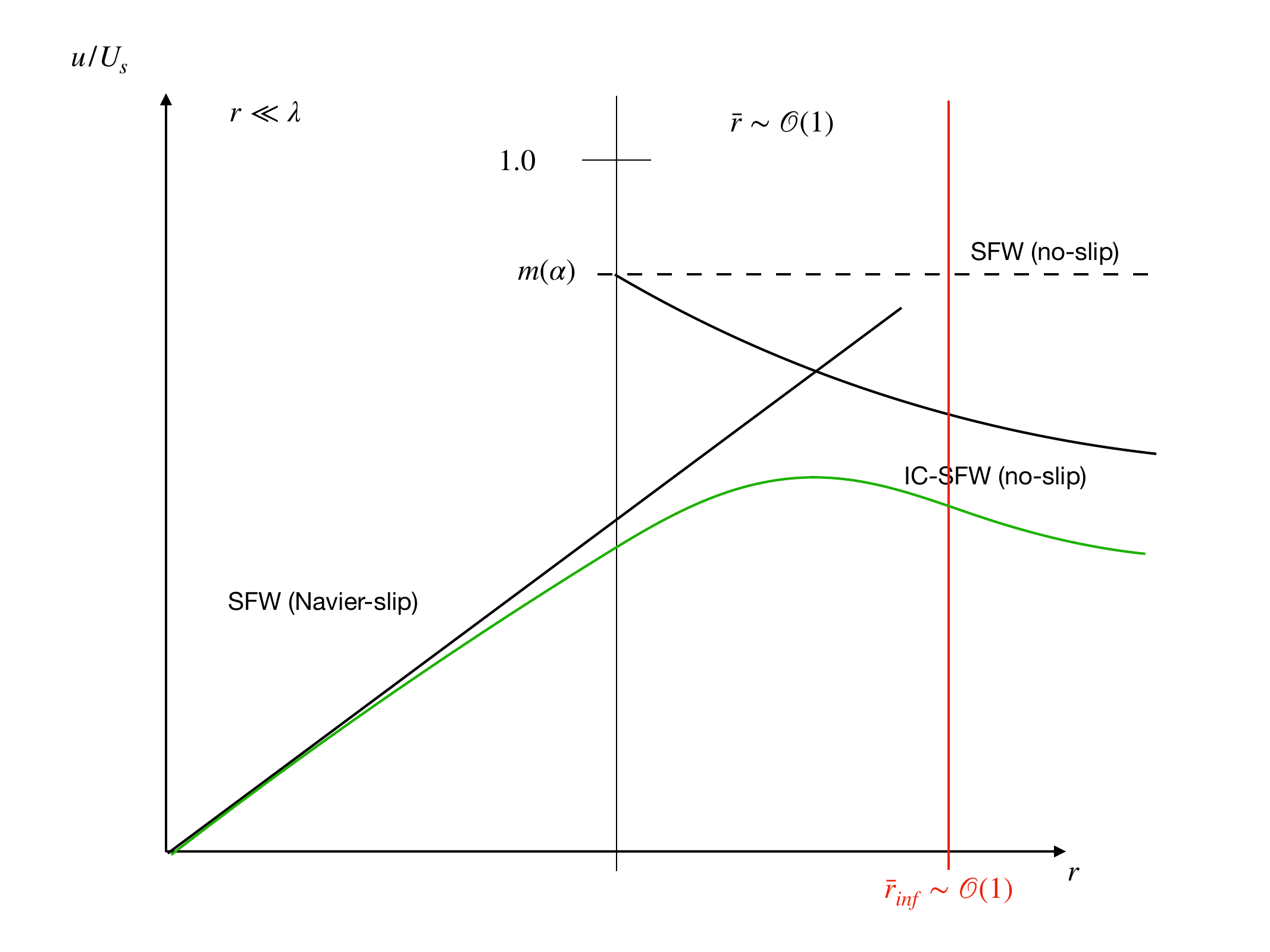}
    \caption{}
    \label{fig:Schematic_IC_SFW_b}
\end{subfigure}

\caption{Schematic of the asymptotic behaviour of the velocity along the interface in the vicinity of the contact line in the reference frame of the contact line. $\Bar{r}$ is the local Reynolds number and $\lambda$ is the slip length. The constant $m(\alpha)$ depends on the wedge angle $\alpha$ and varies from $0$ at $\alpha=0$ to $1$ at $\alpha=\pi$. Subfigure (a) shows the asymptotic behaviour expected if the local Reynolds number based on the inflection point is small enough. Subfigure (b) shows the behaviour if the local Reynolds number based on inflection-point distance is not small enough, which is the case in the present work. This causes us to scale the velocity profiles in Figure~\ref{fig:velocity_iface_FINAL}(b) to $u_{max}$.}
\label{fig:Schemetic_IC_SFW_asymptotics}
\end{figure}

The analytical IC-SFW solution of \citet{Varma} is available only up to the critical angle of $\sim 0.715\pi$, that is around $130^{\circ}$. Hence beyond the critical angle, we use the numerical wedge solution described in \S\ref{sec:velocity_along_iface}.

The effect of adding inertia causes an increase in velocity along the interface that is otherwise absent in the SFW solution. This increase is up to the Stokes-flow value for a given wedge angle. In Figure \ref{fig:basilisk_IC-SFW_all_angles}(a) we show the variation of the velocity along the interface as obtained from our numerical IC-SFW solution. Extracting the limiting value of the velocity in our IC-SFW simulation, we compare it with the theoretically obtained SFW value in Figure \ref{fig:basilisk_IC-SFW_all_angles}(b) for various wedge angles. We see good agreement as we refine the grid.

In the schematic Figure \ref{fig:Schemetic_IC_SFW_asymptotics} we show how the asymptotics are expected to work. The IC-SFW causes velocity to increase up to the SFW value, which depends on the wedge angle. Finally, it decreases to zero as we enter the slip region. Note that $U_{S}$ is the substrate speed, which is also the speed of a fluid particle at the fluid-solid interface under the no-slip condition. In an ideal case, where the slip length is small enough to yield an inflection point close to the contact line such that the local Reynolds number based on the inflection-point distance satisfies $\bar{r}_{inf} \ll 1$, we expect the velocity along the interface to behave like Figure \ref{fig:Schematic_IC_SFW_a}. That is, we expect the velocity along the interface predicted by the IC-SFW solution to exactly match the full curtain region when $\bar{r} \sim \mathcal{O}(1)$. However, in our case, the inflection-point location yields $\bar{r}_{inf} \sim \mathcal{O}(1)$, hence we expect a behaviour like Figure \ref{fig:Schematic_IC_SFW_b}. That is, the velocity along the interface qualitatively behaves similarly to that predicted by the IC-SFW solution, but requires scaling down due to the interference of the slip region. This encourages further work with smaller slip lengths. One could thus imagine the inflection-point distance as an indicator of how far slip-length effects extend.
\section{Additional resolved-slip diagnostics}
\label{app:resolved_slip_diagnostics}
\begin{figure}
    \centering
    \includegraphics[width=0.9\linewidth]{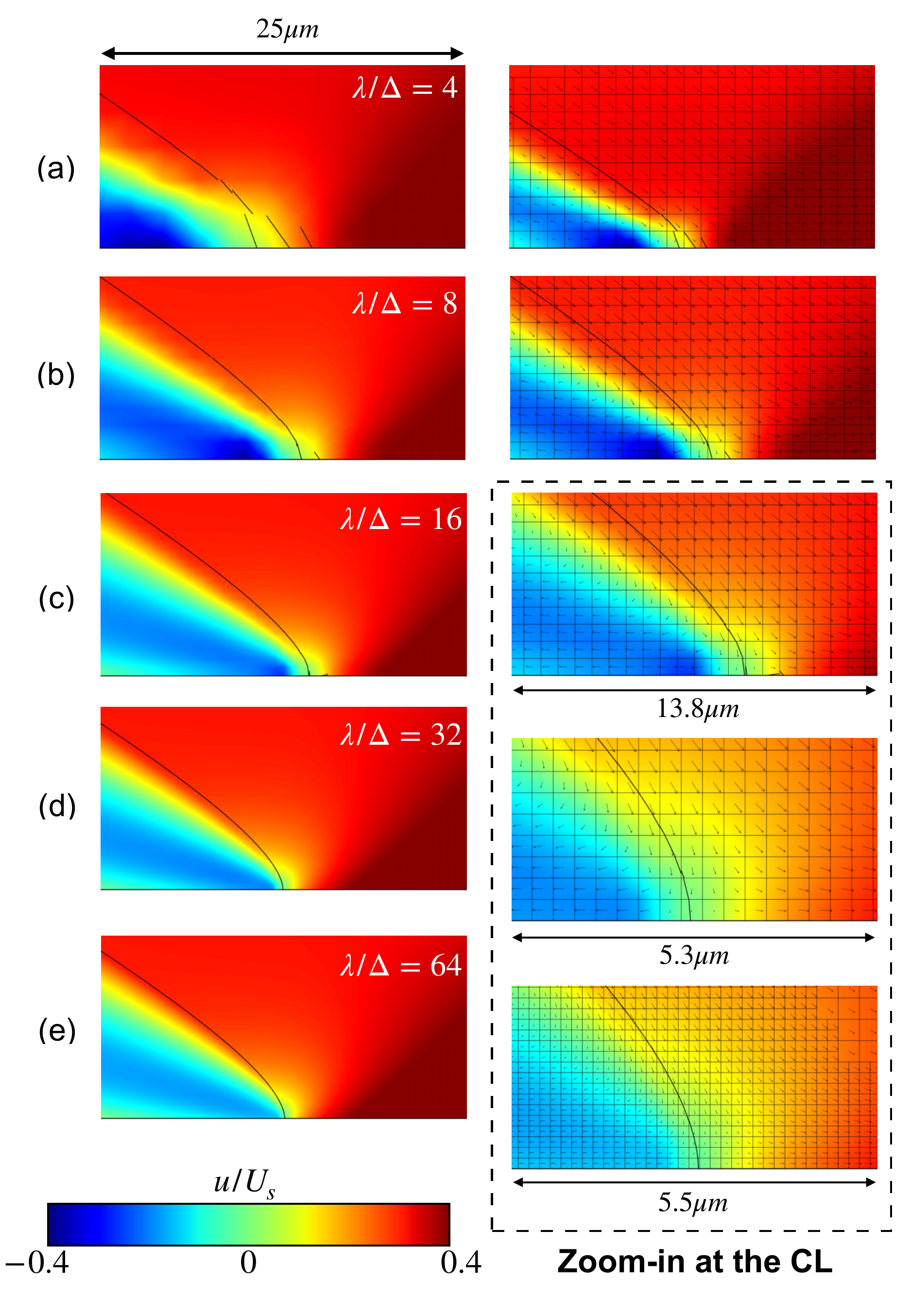}
    \caption{Zoom in near the contact line for $\Reyn = 20$ and $\Ca=0.7$. The horizontal velocity colors the background and the velocity vectors are shown. The resolution in terms of grid size per slip length is (a) $\frac{\lambda}{\Delta} = 4$, (b) $\frac{\lambda}{\Delta} = 8$, (c) $\frac{\lambda}{\Delta} = 16$, (d) $\frac{\lambda}{\Delta} = 32$, and (e) $\frac{\lambda}{\Delta} = 64$. At maximum grid refinement, the grid size is 160 nm. }
    \label{fig:cl_zoom_lvl}
\end{figure}
Figure~\ref{fig:cl_zoom_lvl} shows the progressive improvement of the VoF reconstruction and the local flow field near the contact line as the slip region is increasingly resolved due to mesh refinement.

The main text establishes the logarithmic curvature singularity through Figure~\ref{fig:log_kappa_divergence}. We now provide additional numerical diagnostics showing how this singular behaviour looks with the mesh refinement.
Although the curvature diverges at the contact line, the interface angle at the wall still converges to the imposed value.
In other words, the interface remains locally tangent to the wall with the prescribed contact angle, while the rate at which that angle changes becomes unbounded as the contact line is approached.

\begin{figure}
\centering
\begin{subfigure}[b]{0.49\textwidth}
\centering
\includegraphics[width=\textwidth]{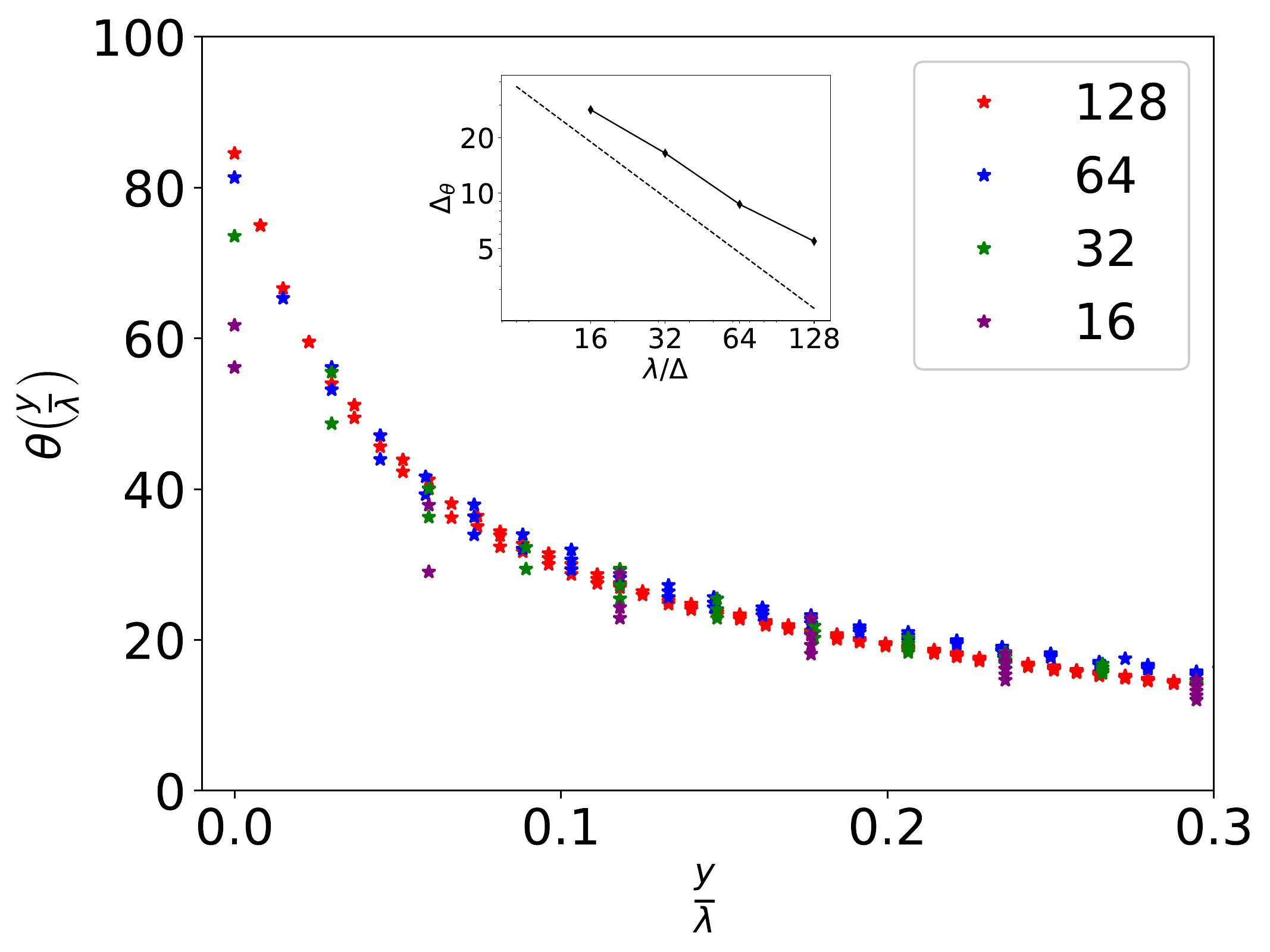}
\caption{}
\label{fig:covergence_angle}
\end{subfigure}
\hfill
\begin{subfigure}[b]{0.49\textwidth}
\centering
\includegraphics[width=\textwidth]{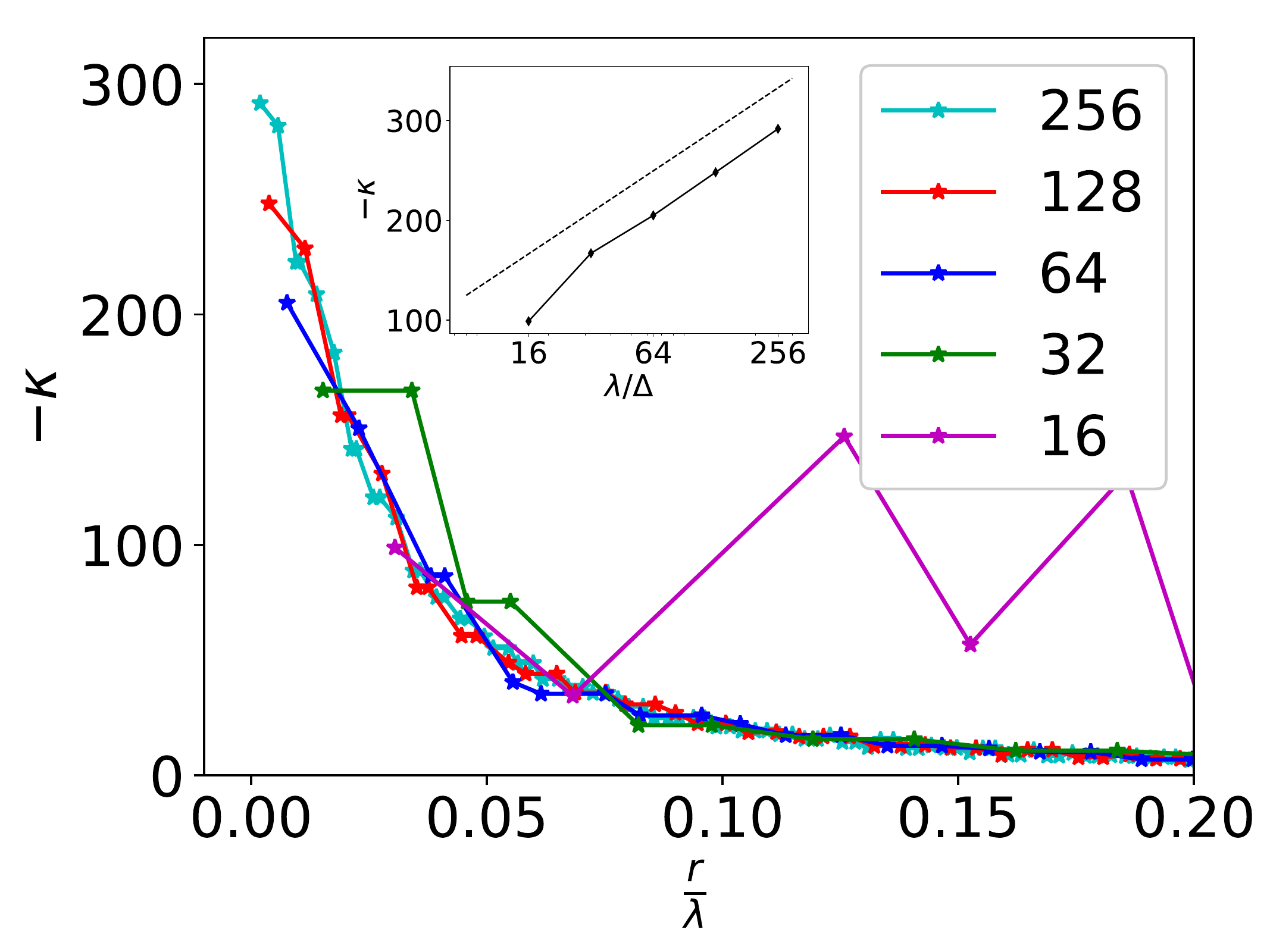}
\caption{}
\label{fig:convergence_kappa}
\end{subfigure}
\caption{(a) Angle or local slope as a function of the vertical distance from the contact line. The inset shows the value of the extracted contact angle in the contact-line cell. (b) Curvature as a function of radial distance from the contact line. The inset shows the extracted curvature from the contact-line cell. Both plots are for $\Reyn=20$ and $\Ca=1.8$. The legend represents the number of grid points per slip length in both cases.}
\label{fig:convergence_kappa_angle}
\end{figure}

Figure~\ref{fig:convergence_kappa_angle} makes this structure more explicit. In Figure~\ref{fig:covergence_angle}, the extracted wall angle converges to the imposed value of $90^\circ$ as the grid is refined. However, the slope profile becomes increasingly steep in the immediate vicinity of the contact line. Equivalently, the derivative of the angle, and hence the curvature, grows without bound. This is shown directly in Figure~\ref{fig:convergence_kappa}, where the curvature extracted closer and closer to the contact line increases systematically with refinement.

This behaviour is the numerical signature of a logarithmic curvature singularity. The singularity is weak (integrable), so the interface remains locally flat at the contact line in the sense of its tangent direction, yet the second derivative diverges. Thus, the contact line may still be viewed locally as having the imposed contact angle, even though the curvature at that same point is unbounded. This is precisely the structure demonstrated in the main text through Figure~\ref{fig:log_kappa_divergence}.

\section{Effect of the microscopic grid-scale contact angle on the large-scale angle in the heel configuration}
\label{app:micro_macro}

\begin{figure}
\centering
\begin{subfigure}[b]{0.49\textwidth}
\centering
\includegraphics[width=\textwidth]{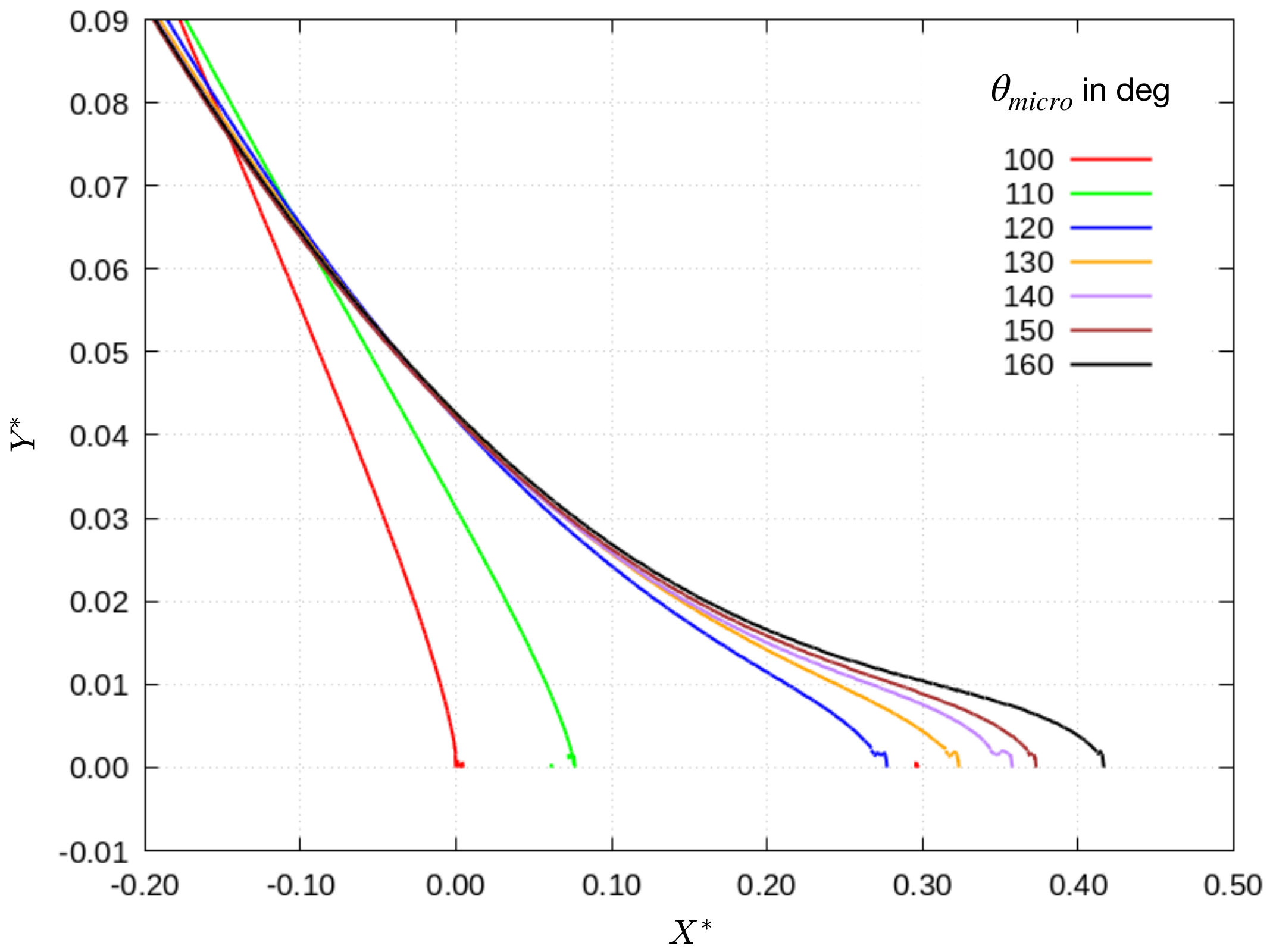}
\caption{}
\label{fig:angle_vary_iface}
\end{subfigure}
\hfill
\begin{subfigure}[b]{0.49\textwidth}
\centering
\includegraphics[width=\textwidth]{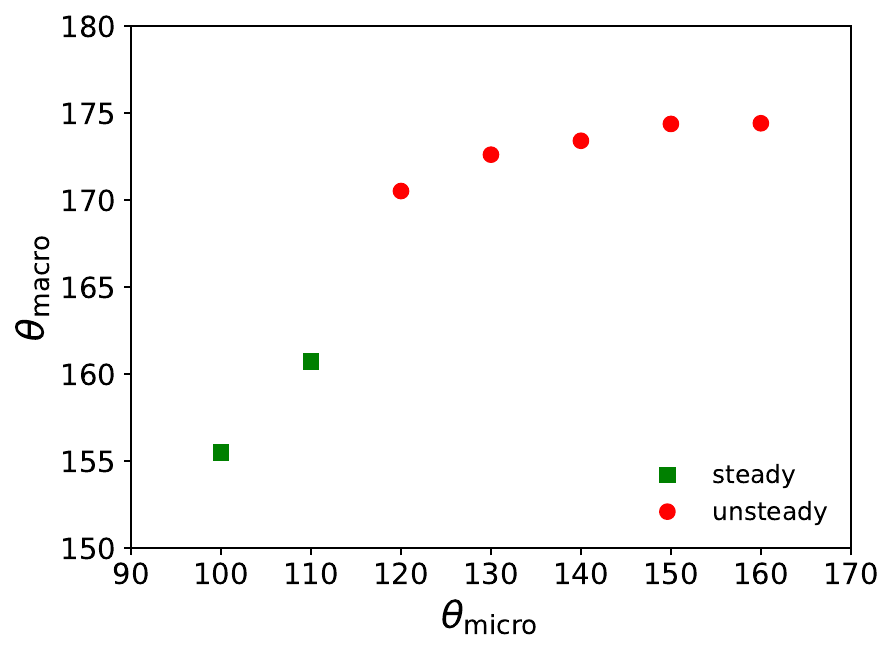}
\caption{}
\label{fig:angle_vary_macro}
\end{subfigure}
\caption{(a) Interface shapes near the contact line for the heel configuration of the reduced model at $\Reyn=30$ and $\Ca=0.7$, for different imposed grid-scale contact angles $\theta_{\mathrm{micro}}$. Only the cases $\theta_{\mathrm{micro}}=100^\circ$ and $110^\circ$ reach steady state; the remaining cases are unsteady. The axes are non-dimensionalised by the coated film thickness, and the resolution is $\lambda/\Delta=16$. (b) Effect of the imposed grid-scale contact angle on the apparent macroscopic angle for the heel configuration of the reduced model at $\Reyn=30$ and $\Ca=0.7$. Here $\theta_{\mathrm{micro}}$ denotes the contact angle imposed at the grid scale through the VoF height-function reconstruction, while $\theta_{\mathrm{macro}}$ is the apparent angle measured at a distance of $20\,\mu\mathrm{m}$ from the contact line. Green squares indicate cases for which a steady-state solution is obtained, whereas red circles indicate unsteady cases that evolve towards air entrainment. The resolution is $\lambda/\Delta=16$.}
\label{fig:angle_vary}
\end{figure}

We examine whether the macroscopic contact angle measured a few tens of microns from the contact line is controlled by the imposed microscopic contact angle. This also clarifies why the free-surface simulations of \citet{Wilson_Yulli} did not reproduce the experimentally observed interface bending of \citet{Blake_1999}.

We consider a heel configuration ($\Reyn=30$, $\Ca=0.7$) and restart from a steady solution obtained with a microscopic angle of $90^\circ$. The microscopic angle is then varied between $100^\circ$ and $160^\circ$. No steady solution exists for imposed angles above $120^\circ$. Figure~\ref{fig:angle_vary_iface} shows the interface shapes after relaxation, where only the $100^\circ$ and $110^\circ$ cases reach steady state.

Figure~\ref{fig:angle_vary_macro} reports the apparent angle measured at $20\,\mu$m from the contact line. When a steady solution exists, the macroscopic angle varies weakly (about $5^\circ$--$7^\circ$) despite large changes in the microscopic angle. This variation is smaller than the experimental uncertainty reported by \citet{Blake_1999}. In contrast, when no steady state exists, the apparent angle approaches $180^\circ$ and film entrainment occurs. Wetting failure is therefore qualitatively different from a steady coating solution rather than a continuation of it.

The free-surface simulations of \citet{Wilson_Yulli} used a microscopic contact angle of $165^\circ$. This directly places them in an unstable regime for the physical two-phase system. However, in a single-phase free-surface formulation with Navier slip, steady solutions exist for arbitrarily large $\Ca$ and imposed angles. As a result, the apparent angle remains close to $180^\circ$ and varies only weakly, leading to a discrepancy with the experimentally measured angle at tens of microns from the contact line.
\section{The DNS chronology of the liquid curtain}
\label{appex:chronology_liquid_curtain}
\begin{figure}
\centerline{\includegraphics[width=0.75\textwidth]{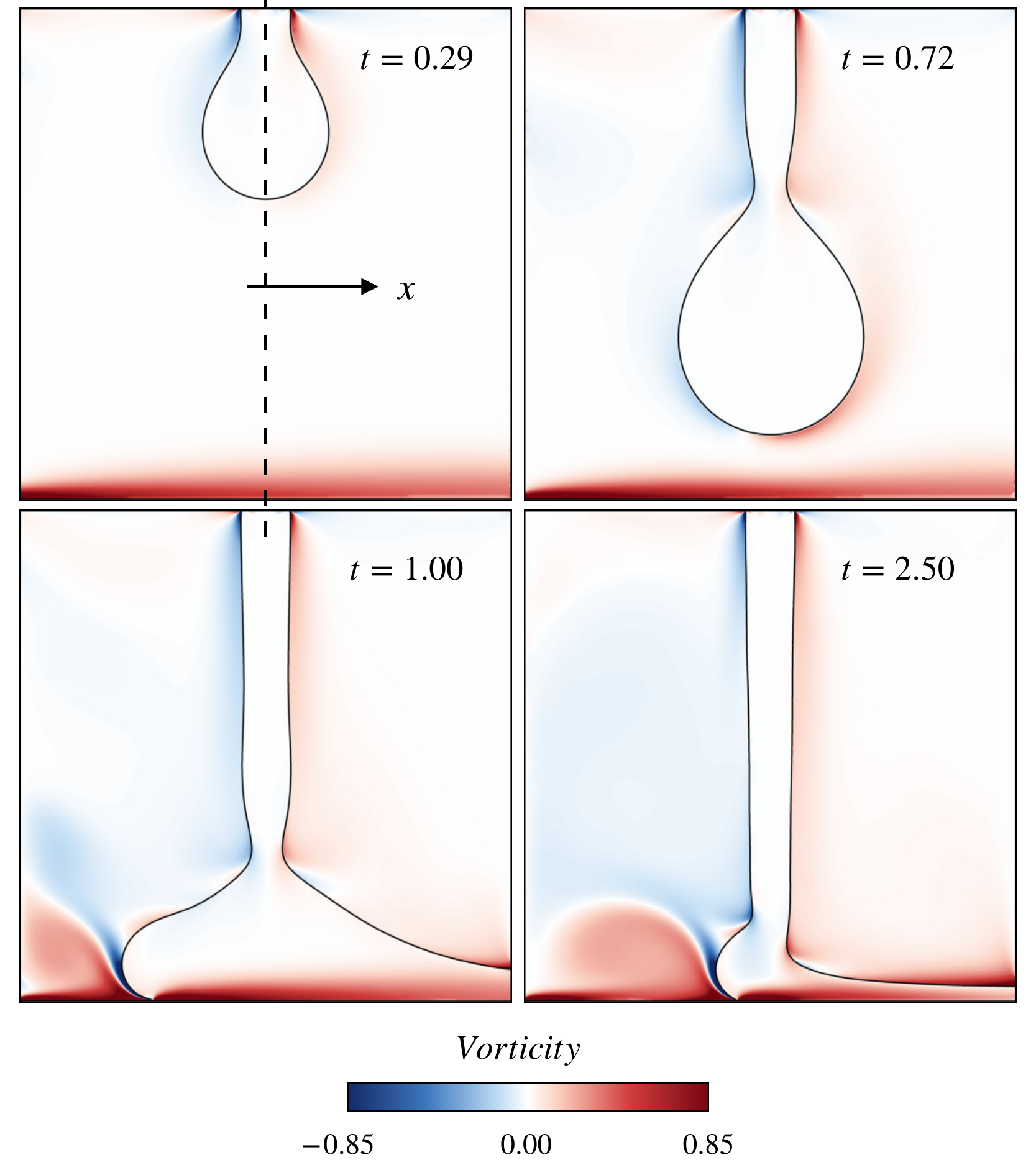}}
\caption{Chronology of the liquid curtain relaxing to a steady state for $\Reyn = 20$ and $\Ca = 0.7$ for the reduced model (section \ref{subsec:compare_exp_Liu}). The plate is being pulled to the right and we are in the laboratory frame of reference. The time is scaled by the free-fall timescale, $T = \sqrt{2h_c/g}$, which is $45\,\mathrm{ms}$, and the background is colored by the non-dimensional vorticity field.}
\label{fig:chronology_curtain}
\end{figure}

Figure \ref{fig:chronology_curtain} shows a chronology of reaching the steady state where the liquid curtain falls and a contact line is formed when the liquid impinges on the substrate. The liquid then spreads on the substrate due to inertia and the contact line moves towards the left in the laboratory frame of reference, reaches a maximum displacement, and then moves towards the right, in the direction of the solid substrate velocity, and reaches a steady-state position. Note that the transient dynamics are out of the scope of this paper and we restrict attention to the steady-state description only.
\FloatBarrier
\begingroup
\section{Construction and verification of the outer-interface arcs}
\label{app:outer_shape_methods}
The three cases are $(\Reyn,\Ca)=(30,0.7)$, $(20,0.7)$ and $(20,1.8)$, all with slip length $\lambda=10\,\mu\mathrm{m}$.

The local Young--Laplace curve is obtained by integrating
\begin{equation}
 \frac{dX}{ds}=\cos\theta,\qquad
 \frac{dY}{ds}=\sin\theta,\qquad
 \frac{d\theta}{ds}=\kappa_T-G(Y-Y_T),
 \label{eq:outer_YL_arc}
\end{equation}
from the measured turn $(X_T,Y_T,\theta_T)=(X_T,Y_T,\pi/2)$, with $\kappa_T$ taken from the simulated interface and $G$ defined in~\eqref{eq:outer_YL_kappa}. Setting $G=0$ gives the circular arc with the same turning-point data.

The curvature $\kappa_T$ is obtained from a locally smoothed reconstruction of the simulated interface around $T$. Halving or doubling the reconstruction width changes $\kappa_T$ by less than $0.6\%$ in all three cases.

Both models are compared on the same interval in each case, beyond the inner inflection transition, as listed in table~\ref{tab:outer_YL_errors}. For the simulated interface $\boldsymbol{r}(s)$, analytical arc $\boldsymbol{R}(\xi)$ and interval length $L_w=s_1-s_0$, the positional error is
\begin{equation}
 E=\left[\frac{1}{L_w}
 \int_{s_0}^{s_1}\min_{\xi}
 \left|\boldsymbol{r}(s)-\boldsymbol{R}(\xi)\right|^2\,ds
 \right]^{1/2}.
 \label{eq:outer_shape_error}
\end{equation}
Here $E$ is the arclength-weighted RMS shortest distance to the analytical arc.

\begin{table}
 \centering
 \begin{tabular}{rrrrrr}
 $\Reyn$ & $\Ca$ & $[Y_{\min},Y_{\max}]$ & $\kappa_T$ & $E_g$ & $E_0$\\
 & & $(\mathrm{mm})$ & $(\mathrm{mm}^{-1})$ & $(\mu\mathrm{m})$ & $(\mu\mathrm{m})$\\
 \hline
 30 & 0.7 & $[0.063,1.634]$ & 0.915 & 7.54 & 16.91\\
 20 & 0.7 & $[0.049,1.086]$ & 1.530 & 5.09 & 5.98\\
 20 & 1.8 & $[0.023,0.600]$ & 1.792 & 16.27 & 16.70
 \end{tabular}
 \caption{Comparison intervals, measured turning-point curvature and RMS positional errors. $E_g$ uses physical gravity; $E_0$ uses the circular limit.}
 \label{tab:outer_YL_errors}
\end{table}
\par
\endgroup

\FloatBarrier
\bibliographystyle{jfm}
\bibliography{references}
\end{document}